\documentclass[lettersize,journal]{IEEEtran}
\IEEEoverridecommandlockouts

\usepackage{cite}
\usepackage{comment}

\usepackage{amsmath,amssymb,amsfonts}
\usepackage{algorithm}
\usepackage[noend]{algpseudocode}
\usepackage{graphicx}
\usepackage{textcomp}
\usepackage[dvipsnames]{xcolor}
\usepackage{tabularx}

\usepackage{mathtools}
\usepackage{mwe}  
\usepackage{siunitx}
\usepackage{booktabs}
\let\mc\multicolumn

\usepackage{macros}

\usepackage[font=small,labelfont=bf,tableposition=top]{caption}
\DeclareCaptionLabelFormat{andtable}{#1~#2  \&  \tablename~\thetable}

\def\BibTeX{{\rm B\kern-.05em{\sc i\kern-.025em b}\kern-.08em
    T\kern-.1667em\lower.7ex\hbox{E}\kern-.125emX}}
\begin{document}

\title{Data-Driven Estimation of Failure Probabilities in Correlated Structure-Preserving Stochastic Power System Models
\thanks{This material is based upon work supported by the
	U.S. Department of Energy, Office of Science, Advanced Scientific
	Computing Research under Contract DE-AC02-06CH11357. The work reported in this paper has been partly supported by
the U.S. Department of Energy Advanced Grid Modeling Program.}
}

\author{Hongli Zhao$^{\dagger\S}$, Tyler E. Maltba$^{\ddagger}$, D. Adrian Maldonado$^{\S}$, Emil Constantinescu$^{\S}$, Mihai Anitescu$^{\S}$
\\
$^{\S}$ Mathematics and Computer Science Division, Argonne National Laboratory, Lemont, IL, USA \\

$\ddagger$ Theoretical Division, Los Alamos National Laboratory, Los Alamos, NM, USA \\

$^{\dagger}$ Department of Statistics, University of Chicago, Chicago, IL, USA

}

\thispagestyle{plain}
\pagestyle{plain}
\maketitle

\begin{abstract} We propose a data-driven approach for propagating uncertainty in stochastic power grid simulations and apply it to the estimation of transmission line failure probabilities. A reduced-order equation governing the evolution of the observed line energy probability density function is derived from the Fokker--Planck equation of the full-order continuous Markov process. Our method consists of estimates produced by numerically integrating this reduced-order equation. Numerical experiments for scalar- and vector-valued energy functions are conducted using the classical multimachine model under spatiotemporally correlated noise perturbation. The method demonstrates a more sample-efficient approach for computing probabilities of tail events when compared with kernel density estimation. Moreover, it produces vastly more accurate estimates of joint event occurrence when compared with independent models. 
\end{abstract}


\section{Introduction} Recent blackout events and fires caused by power lines have continued to raise awareness of the need for designing robust circuit systems and devising mitigation strategies when catastrophic events occur \cite{Lenzi2023}. Uncertainties in peak power demands, operative states of components, load, and seasonal factors contribute to considerable difficulties in modeling line failures \cite{10.1063/1.2737822,8735917}. 

Failure events are typically studied through traversal/clustering of time-dependent graphs \cite{8362273} or by solving differential systems that incorporate transient/line-removal dynamics \cite{FRASCA2021111460}. Particularly for large cases, the first class of approaches is restricted to predicting local behavior that may not yield accurate results corresponding to real-world observations \cite{7501508}. On the other hand, extensive simulations are needed in order to obtain statistical information of failure distributions as tail events, making the dynamical systems approach computationally intensive. In this work, we build upon the systematic approach of \cite{6547228} to incorporate uncertainty in a structure-preserving dynamics model by including the Ornstein--Uhlenbeck (OU) process as load stochastic fluctuations. In the absence of approximations, resolving the full probability profile of all random states suffers from the curse of dimensionality. Although reduction methods such as the method of moments (MoM) \cite{8586023} and Karhunen--Lo\`eve (KL) expansions \cite{9136799} exist, they are known to underperform standard Monte Carlo simulations and kernel density estimation (KDE) for realistic power system models. In particular, the MoM requires state data to be nearly Gaussian, while KL expansions degrade when the noise input exhibits a short correlation time scale \cite{MALTBA201887}.

To address this need, instead of approximating the full joint probability density, we exploit the fact that in the model of \cite{6547228} the probability of failure of a transmission line depends on a scalar function that can be efficiently evaluated. Subsequently, we directly propagate its probability density function (PDF) by expressing it as a solution of a low-dimensional partial differential equation (PDE) that we solve numerically. This reduced-order formulation allows us to simultaneously consider the joint density of multiple transmission lines, whose conditional probabilistic structure is typically difficult to simulate using kinetic Monte Carlo methods \cite{roth2019kinetic}. Furthermore, the method is data-driven, where measurements from system states can be incorporated into estimating unknown terms in the PDE using regression methods. The resulting PDE is a linear equation that is readily solved with standard finite volume schemes, with comparable accuracy to plug-in Gaussian KDE, while using significantly fewer samples from the law of the underlying stochastic process. We also note that the previous approach in \cite{roth2019kinetic}, while being more sample-efficient, computed only the large deviation (low-temperature limit) approximation of the failure probability whereas this approach with enough samples will converge to the true probability of failure. 

The paper is organized as follows. Section~\ref{sec:model} introduces the general power system model along with the correlated noise process, which is jointly considered as a diffusion process admitting a Fokker--Planck equation (FPE). In Section~\ref{sec:method} we discuss the derivation of the reduced-order equation by integrating over extraneous variables in the FPE, allowing one to formulate the density of arbitrary quantities of interest. In addition, we discuss estimation of unknown terms in the PDE as a regression problem. Effectiveness of the method is tested in Section~\ref{sec:cases} on the classical multimachine model with an emphasis on use cases of computing tail event probabilities and correlation. In Section~\ref{sec:conclude} we summarize our work and discuss possible directions for extension.

\section{Stochastic Power System Model} \label{sec:model} A stochastic power system model is generally described by a set of index-1 governing differential algebraic equations:
\begin{equation} \label{eqn:dae-system}
\begin{split}
    \dot{\boldsymbol{x}}(t) &= \boldsymbol{f}(\boldsymbol{x}(t),\boldsymbol{y}(t); \boldsymbol{\eta}(t)) 
    \\
    \boldsymbol{0} &= \boldsymbol{g}(\boldsymbol{x}(t), \boldsymbol{y}(t); \boldsymbol{\eta}(t)),
\end{split}
\end{equation} where $\boldsymbol{x}(t)$ are the system states at time $t$ and $\boldsymbol{y}(t)$ accounts for any auxiliary or algebraic variables. Furthermore, to account for random fluctuations in the states, a noise process $\boldsymbol{\eta}(t)$ is introduced with the following general form~\cite{6547228}:
\begin{equation} \label{eqn:correlated-wiener}
\begin{split}
    \dot{\boldsymbol{\eta}}(t)   & = \boldsymbol{a}(\boldsymbol{\eta}(t)) + \boldsymbol{b}(\boldsymbol{\eta}(t)) \odot \dot{\boldsymbol{\zeta}}(t) \\
    \dot{\boldsymbol{\zeta}}(t)  & = \boldsymbol{C}\cdot \frac{d\boldsymbol{w}(t)}{dt} ,
\end{split}
\end{equation} where $\boldsymbol{w}(t)$ denotes the multidimensional standard Wiener process in $\mathbb{R}^n$, such that $d\boldsymbol{w}/dt$ is formally interpreted as white noise. In essence, the second equation of (\ref{eqn:correlated-wiener}) defines a set of correlated Wiener processes with correlation matrix $\boldsymbol{R}:=\boldsymbol{C}\boldsymbol{C}^T$. We assume Lipschitz continuity of $\boldsymbol{a}, \boldsymbol{b}$ in (\ref{eqn:correlated-wiener}), such that (\ref{eqn:dae-system}) admits the following form of diffusion processes, where we define $\boldsymbol{z}(t) := [\boldsymbol{x}^T(t), \boldsymbol{y}^T(t), \boldsymbol{\eta}^T(t)]^T\in \mathbb{R}^N$, with $N\gg n$:
\begin{equation} \label{eqn:general-form-sde}
\begin{split}
    d\boldsymbol{z}(t) &= \boldsymbol{\mu}(\boldsymbol{z}(t),t)dt + \boldsymbol{\sigma}(\boldsymbol{z}(t),t)d\boldsymbol{W}(t) \\
    \boldsymbol{z}(0) &\sim f_{\boldsymbol{z}_0},
\end{split}
\end{equation} where $\boldsymbol{W}(t)$ is the $\rr^N$-valued standard Wiener process. The initial condition $\boldsymbol{z}_0$ may be either random with probability density $f_{\boldsymbol{z}_0}$ or deterministic, in which case $f_{\boldsymbol{z}_0}\equiv\boldsymbol{\delta}_{\boldsymbol{Z}^*}$ is the (multidimensional) Dirac delta distribution at a fixed point $\boldsymbol{Z}^*$. Furthermore, let $\boldsymbol{\mu}_t \equiv \boldsymbol{\mu}(\boldsymbol{z}(t),t): \rr^N\times\rr^{+}\rightarrow \rr^N$ define the vector of drifts, and let $\boldsymbol{\sigma}_t \equiv \boldsymbol{\sigma}(\boldsymbol{z}(t),t): \rr^N\times\rr^{+}\rightarrow\rr^{N\times N}$ define the diffusion matrix. The velocity and noise terms of (\ref{eqn:general-form-sde}) are assumed to satisfy sufficient regularity, which guarantee the existence of solutions whose paths are almost surely continuous.

The It\^o process (\ref{eqn:general-form-sde}) admits a full-order Fokker--Planck equation \cite{Risken1996}:
\begin{equation}\label{eqn:general-fpe}
    \begin{split}
     \frac{\partial f_{\boldsymbol{z}}}{\partial t} + \sum_{i=1}^N\frac{\partial}{\partial Z_i}(\mu_i(\boldsymbol{Z}, t)f_{\boldsymbol{z}}) &= \sum_{i,j=1}^N\frac{\partial^2}{\partial Z_i\partial Z_j} (\mathcal{D}_{ij}(\boldsymbol{Z},t)f_{\boldsymbol{z}}) \\
     f_{\boldsymbol{z}}(\boldsymbol{Z},0) &= f_{\boldsymbol{z}_0}(\boldsymbol{Z}) ,
    \end{split}
\end{equation} The advection and diffusion coefficients of Equation (\ref{eqn:general-fpe}) are assumed to decay sufficiently fast at $\pm\infty$ so that vanishing boundary conditions can be employed. Furthermore, we denote $\mathcal{D} := \frac12\boldsymbol{\sigma}_t\boldsymbol{\sigma}_t^T \in \rr^{N\times N}$. When drift and diffusion are time-dependent, it is generally not the case that (\ref{eqn:general-fpe}) possesses a stationary solution. We refer the reader to \cite{old-stationary-paper} for specific investigated cases where an analytic derivation is possible. 

\section{Proposed Method} \label{sec:method} To estimate probabilities of certain events defined using a forward model, one may either simulate Monte Carlo trajectories of (\ref{eqn:general-form-sde}) and post-process with KDE, or solve (\ref{eqn:general-fpe}) and perform numerical integration on a computational mesh. Both approaches suffer from the curse of dimensionality when $N$ becomes large \cite{091533b0-fe38-3eea-b846-fd10afc72174,Freidlin2012}. Instead of seeking to resolve the full-state dynamics, it often suffices for practical purposes to observe the evolution of a low-dimensional quantity of interest (QoI) over the (finite) domain of numerical simulations. Therefore, we consider deriving the PDF dynamics for the QoI directly, and solve only a low-dimensional FPE-like equation (referred to as the ``reduced-order PDF equation'' of the QoI). In this section, we present a general extension for formally deriving the PDF equation for QoIs of arbitrary dimensions. We begin by reviewing the derivation for scalar QoIs in the next section.

\subsection{Reduced-Order PDF Equations} \label{sec:1dropdf-derive} We let $u:\rr^{N}\rightarrow\rr$ be a deterministic, second-order continuously differentiable function of the system states, representing a QoI evaluated from the states of system (\ref{eqn:general-form-sde}). Furthermore, let $u(t) \equiv u(\boldsymbol{z}(t))$ denote the state of the QoI at time $t$, and let $U\in \rr$ denote the corresponding phase space variable. Applying It\^{o}'s lemma, we see that the QoI follows an associated process:
\begin{align} \label{eqn:reduced-order-process}
\begin{split}
    du(t) = \big(
        (\nabla_{\boldsymbol{z}}u)^T\boldsymbol{\mu}_t 
    &+
        \frac12\text{tr}(\boldsymbol{\sigma}_t^T(H_{\boldsymbol{z}}u)\boldsymbol{\sigma}_t)
    \big)dt \\
    &+ (\nabla_{\boldsymbol{z}}u)^T\boldsymbol{\sigma}_td\boldsymbol{W}_t ,
\end{split}
\end{align} where we let $\nabla_{\boldsymbol{z}} := [\frac{\partial}{\partial z_1},\ldots, \frac{\partial}{\partial z_N}]^T$. $H_{\boldsymbol{z}}(\cdot)$ denotes the Hessian operator with respect to states $\boldsymbol{z}(t)$. For simplicity of notations, we let $\mu^u$ denote the drift term of the process (\ref{eqn:reduced-order-process}). To derive a deterministic governing equation for the probability density $f_u$, we augment the states by considering the joint vector $\tilde{\boldsymbol{z}}(t):=[\boldsymbol{z}^T(t), u(t)]^T\in\rr^{N+1}$. Similarly, let $\widetilde{\boldsymbol{W}}(t) := [\boldsymbol{W}^T(t), w(t)]^T$ denote an $(N+1)$-dimensional standard Wiener process, with $w(t)$ independent of $\boldsymbol{W}(t)$. We arrive at the following system:
\begin{equation} \label{eqn:augmented-process}
    \begin{split}
         d\tilde{\boldsymbol{z}}(t)
        &= \widetilde{\boldsymbol{\mu}}_tdt + \widetilde{\boldsymbol{\sigma}}_td\widetilde{\boldsymbol{W}}(t)\\
        &= \begin{bmatrix}
            \boldsymbol{\mu}_t \\
            \mu^u
        \end{bmatrix}dt + 
        \begin{bmatrix}
            \boldsymbol{\sigma}_t & \boldsymbol{0} \\
            (\nabla_{\boldsymbol{z}}u)^T\boldsymbol{\sigma}_t & \boldsymbol{0}
            \end{bmatrix}d\widetilde{\boldsymbol{W}}(t).
    \end{split}
\end{equation} Since the function $u$ is deterministic, it does not introduce additional diffusion. Similar to (\ref{eqn:general-fpe}), the density of the augmented process \eqref{eqn:augmented-process} follows the $(N+1)$-dimensional FPE for the joint states with the same form as (\ref{eqn:general-fpe}). After factoring the joint density in terms of marginal and conditional densities, that is, the decomposition $f_{\tilde{\boldsymbol{z}}} = f_{\boldsymbol{z}|u}\cdot f_u$, we marginalize over the state space of $\boldsymbol{z}$ and arrive at the following reduced-order PDF equation in $u$ only:
\begin{equation} \label{eqn:reduced-order-pdf-equation}
\begin{split}
    \frac{\partial f_u}{\partial t} &+ \frac{\partial}{\partial U}\big(
        \expect{
            \mu^u \,|\, u(t) = U
        }f_u
    \big) \\ 
    & =
    \frac{\partial^2}{\partial U^2}\big(
        \expect{
            \mathcal{D}^u \,|\, u(t)= U
        }f_u
    \big) \\
    f_{u}(&U,0) = f_{u_0}(U),
\end{split}
\end{equation} where $f_{u_0}$ denotes the initial density of the QoI, which may be either analytically known or estimated from a KDE using evaluated samples, and
\begin{equation}\label{eqn:reduced-order-drift}
   \mu^u := (\nabla_{\boldsymbol{z}}u)^T\boldsymbol{\mu}_t + 
    \frac12\text{tr}(\boldsymbol{\sigma}_t^T(H_{\boldsymbol{z}}u)\boldsymbol{\sigma}_t)
\end{equation} and
\begin{equation} \label{eqn:reduced-order-diffusion}
\begin{split}
    \mathcal{D}^u 
    &:= (\nabla_{\boldsymbol{z}}u)^T\mathcal{D}(\nabla_{\boldsymbol{z}}u) = \frac12[\widetilde{\boldsymbol{\sigma}}_t\widetilde{\boldsymbol{\sigma}}_t^T]_{(N+1)(N+1)}.
\end{split}
\end{equation} 

As an illustration, we consider the special case where the QoI denotes a projection onto a specific coordinate of $\boldsymbol{z}$.

\emph{Example: } (Coordinate Projection) Let $u(\boldsymbol{z}(t)) = z_k(t)$ for some fixed $1\le k\le N$.  We have
\begin{equation} \label{eqn:coord-projection-qoi}
    \nabla_{\boldsymbol{z}}u \equiv \boldsymbol{e}_k, \quad H_{\boldsymbol{z}}u \equiv \boldsymbol{0},
\end{equation} 
where $\boldsymbol{e}_k := [0, \ldots, 0, 1, 0, \ldots, 0]^T$ is the $k$th standard basis vector in $\rr^N$. Substituting (\ref{eqn:coord-projection-qoi}) in the derivations of (\ref{eqn:reduced-order-drift}) and (\ref{eqn:reduced-order-diffusion}), we obtain
\begin{equation}
     \mu^u \equiv \mu_k, \quad
     \mathcal{D}^u \equiv \mathcal{D}_{kk},
\end{equation} where $\mu_k$ is the $k$th coordinate of the drift vector $\boldsymbol{\mu}_t$, such that the PDF equation becomes
\begin{equation}
\begin{split}
    \frac{\partial f_{z_k}}{\partial t} 
    &+ \frac{\partial}{\partial Z_k}\big(
        \expect{\mu_k(\boldsymbol{z}(t),t)\,|\, z_k(t) = Z_k}f_{z_k}
    \big) \\
    &= \frac{\partial^2}{\partial Z_k^2}\big(
        \expect{\mathcal{D}_{kk}(\boldsymbol{z}(t),t) \,|\, z_k(t) = Z_k}f_{z_k}
    \big),
\end{split}
\end{equation} which is consistent with the results presented in \cite{Brennan_2018,maltba2022learning} for single components of the state vector.

\subsection{Joint Reduced-Order PDF Equations} \label{sec:joint-ropdf-equations} In applications where conditional events are involved, it is helpful to consider the joint probabilistic structure of vector-valued QoIs. The derivations presented in Section~\ref{sec:1dropdf-derive} can be readily extended to multiple dimensions. To do so, we define a multivariate mapping with second-order continuously differentiable components, $\boldsymbol{u}(\boldsymbol{z}(t)) = [u_1(\boldsymbol{z}(t)), \ldots, u_{N_{\text{R}}}(\boldsymbol{z}(t))]^T$, with $1 < N_{\text{R}} \ll N$. Each component $u_j, 1\le j\le N_{\text{R}}$ of the vector-valued mapping $\boldsymbol{u}$ is understood as a mapping from $\rr^N$ to $\rr$. We apply the multivariate extension of It\^{o}'s lemma and observe that the vector of QoIs, $\boldsymbol{u}(t) \equiv \boldsymbol{u}(\boldsymbol{z}(t))$, follows a diffusion process using the same derivation in (\ref{eqn:reduced-order-process}) for each component:
\begin{equation} \label{eqn:joint-reduced-order-sde}
    d\boldsymbol{u}(t) = \big(\boldsymbol{J}_{\boldsymbol{u}}\boldsymbol{\mu}_t + \boldsymbol{h}_{\boldsymbol{u}}
    \big)dt + \boldsymbol{J}_{\boldsymbol{u}}\boldsymbol{\sigma}_td\boldsymbol{W}_t,
\end{equation} where $\boldsymbol{J}_{\boldsymbol{u}}\in\rr^{N_{\text{R}}\times N}$ is the Jacobian matrix of the vector-valued function $\boldsymbol{u}(\cdot)$ with respect to the states $\boldsymbol{z}(t)$, and
\begin{equation} \label{eqn:define-h-u}
    \boldsymbol{h}_{\boldsymbol{u}} := 
    \frac{1}{2}
    \begin{bmatrix}
        \text{tr}(\boldsymbol{\sigma}_t^T(H_{\boldsymbol{z}}u_1)\boldsymbol{\sigma}_t) \\ 
        \vdots \\
        \text{tr}(\boldsymbol{\sigma}_t^T(H_{\boldsymbol{z}}u_{N_{\text{R}}})\boldsymbol{\sigma}_t) 
    \end{bmatrix}.
\end{equation} 

With a similar argument to (\ref{eqn:reduced-order-pdf-equation}), we arrive at the reduced-order PDF equation joint in multiple QoIs, which is a deterministic PDE of $N_{\text{R}}$-dimensions:
\begin{equation} \label{eqn:m-dim-ropdf}
\begin{split}
    \frac{\partial f_{\boldsymbol{u}}}{\partial t} 
    &+ \sum_{i=1}^{N_{\text{R}}}\frac{\partial}{\partial U_i}(\expect{\mu_i^{\boldsymbol{u}}\,|\,\boldsymbol{u}(t)=\boldsymbol{U}}f_{\boldsymbol{u}}) \\
    &= \sum_{i,j=1}^{N_{\text{R}}}\frac{\partial^2}{\partial U_i\partial U_j}(\expect{\mathcal{D}_{ij}^{\boldsymbol{u}}\,|\,\boldsymbol{u}(t) = \boldsymbol{U}}f_{\boldsymbol{u}}) \\
    f_{\boldsymbol{u}}(&\boldsymbol{U},0) = f_{\boldsymbol{u}_0}(\boldsymbol{U}),
\end{split}
\end{equation} where we defined the vector of drifts for the chosen subset of QoIs:
\begin{equation} \label{eqn:multiple-ropdf-drift}
    {\boldsymbol{\mu}}^{\boldsymbol{u}} = [{\mu}_{1}^{\boldsymbol{u}}, \ldots, {\mu}_{N_{\text{R}}}^{\boldsymbol{u}}]^T := 
    \boldsymbol{J_{u}}\boldsymbol{\mu}_t + \boldsymbol{h_u}
\end{equation} and the augmented matrix similar to that of (\ref{eqn:reduced-order-diffusion}) such that
\begin{equation}
    {\mathcal{D}}^{\boldsymbol{u}} := \boldsymbol{J}_{\boldsymbol{u}}\mathcal{D}\boldsymbol{J}_{\boldsymbol{u}}^T \in \mathbb{R}^{N_{\text{R}}\times N_{\text{R}}}, 
\end{equation} where $\mathcal{D}$ is defined in (\ref{eqn:general-fpe}). In particular, the state-dependent conditional expectations in the reduced-order PDF equation (\ref{eqn:m-dim-ropdf}) are $N_{\text{R}}$-variate functions of the QoIs $\boldsymbol{u}(t)$. 

\subsection{Approximation of Closure Terms} The analytic expressions of the advection and diffusion coefficients in equation (\ref{eqn:m-dim-ropdf}) involve high-dimensional integrals and generally do not admit elementary expressions. Numerical integration or expansion-based approximations (e.g., generalized cumulant expansions) are strongly dependent on the original system dimension $N$ and become intractable for $N\gg1$. 

However, a direct data-driven approximation of the unknown terms is possible by recognizing the estimation of such conditional expectations as a minimization problem under the $L^2$-norm induced by the probability distribution of $\boldsymbol{u}(t)$~\cite{Hastie2009}:
\begin{equation} \label{eqn:l2-regression}
    \begin{split}
        \mathcal{R}(\boldsymbol{U}; t) &= \expect{h(t)\,|\,\boldsymbol{u}(t)=\boldsymbol{U}} \\ &\in \argmin_{g\in L^2(\boldsymbol{u};t)}\mathbb{E}\big[\abs{h(t)-g(\boldsymbol{u}(t))}^2\big] ,
    \end{split}
\end{equation} where $h(t)\equiv h(\boldsymbol{z}(t);\omega)$ is a random field depending on the states $\boldsymbol{z}(t)$ and is homogenized under the conditional law of $\boldsymbol{u}(t)$ as a result of solving (\ref{eqn:l2-regression}). The minimization problem (\ref{eqn:l2-regression}) can be approximately solved by using regression-based methods from simulated trajectories of $h(t)$ and $\boldsymbol{u}(t)$. The estimated coefficients are henceforth referred to as the ``regression functions'' of the reduced-order equations. With the main steps outlined, we summarize the procedure for the reduced-order PDF equation in the following section.

\subsection{Implementation of Reduced-Order PDF Method} The implementation of (\ref{eqn:m-dim-ropdf}) requires a choice of numerical PDE scheme (e.g., finite difference, finite volume, finite element) coupled with a routine for identifying the coefficient terms (\ref{eqn:l2-regression}) (e.g., parametric or nonparametric regression methods). For completeness, we present a practical implementation in Algorithm~\ref{alg:mainalg} based on the finite volume method, which involves an explicit scheme resolving (\ref{eqn:m-dim-ropdf}) and estimation of advection and diffusion coefficients through the general minimization problem (\ref{eqn:l2-regression}).

\begin{algorithm} 
\caption{Data-driven finite volume scheme with adaptive time-stepping. }
\begin{algorithmic}[1]
\item Simulate/measure system (\ref{eqn:dae-system}) to generate Monte Carlo trajectories of $\boldsymbol{z}(t)\in\rr^N$ at discretized time steps $t_1, t_2, \ldots t_{N_t}$. Compute Monte Carlo data for input $\boldsymbol{u}(\boldsymbol{z}(t))$ and ``response variables'' $\boldsymbol{\mu}^{\boldsymbol{u}}(t)$, ${\mathcal{D}}^{\boldsymbol{u}}(t)$. 
\item Estimate initial density function $f_{\boldsymbol{u}_0}$ via kernel density estimation, if not known analytically. 
    \For{$i=1,2,\ldots, N_t$}
        \State Estimate regression functions at time $t_{i-1}$ by approximately solving (\ref{eqn:l2-regression}), yielding
        \begin{equation}
            \begin{split}
                \widehat{\mathcal{R}}_{\mu_i^{\boldsymbol{u}}}(\boldsymbol{U}; t_{i-1}) &\approx \expect{\mu_i^{\boldsymbol{u}}|\boldsymbol{u}(t_{i-1}) = \boldsymbol{U}} \\
                \widehat{\mathcal{R}}_{{\mathcal{D}}_{ij}^{\boldsymbol{u}}}(\boldsymbol{U}; t_{i-1}) &\approx \expect{{\mathcal{D}}_{ij}^{\boldsymbol{u}}|\boldsymbol{u}(t_{i-1}) = \boldsymbol{U}} .
            \end{split}
        \end{equation}
        
        \State Adjust time step by the Courant--Friedrichs--Lewy (CDF) condition for explicit schemes.

        \State Interpolate coefficients $\widehat{\mathcal{R}}_{\mu_i^{\boldsymbol{u}}}, \widehat{\mathcal{R}}_{{\mathcal{D}}_{ij}^{\boldsymbol{u}}}$ in time.
        
        \State Time step the following approximate PDE for $\Delta t$ with an explicit finite volume scheme (e.g., upwind method):
        \begin{equation}
            \begin{split}
                \frac{\partial f_{\boldsymbol{u}}}{\partial t} 
                &+ \sum_{i=1}^{N_{\text{R}}}\frac{\partial}{\partial U_i}(\widehat{\mathcal{R}}_{\mu_i^{\boldsymbol{u}}}f_{\boldsymbol{u}}) =
                \sum_{i,j=1}^{N_{\text{R}}}\frac{\partial^2}{\partial U_i\partial U_j}(\widehat{\mathcal{R}}_{{\mathcal{D}}_{ij}^{\boldsymbol{u}}}f_{\boldsymbol{u}}).
            \end{split}
        \end{equation}
    \EndFor
\item \textbf{end for}
\end{algorithmic} \label{alg:mainalg}
\end{algorithm}

The main source of error in this procedure lies in the estimation and interpolation of advection and diffusion coefficients to the refined temporal grids (as constrained by the CFL condition), where Monte Carlo data is not necessarily available. Apart from interpolating at discrete times, it is also possible to fit a time-varying regression function beforehand and query the advection and diffusion coefficients at arbitrary time points as needed \cite{Zhang_2015}, though at a significantly higher computational overhead. For simplicity, we estimate the regression functions at discrete times and linearly interpolate them to required refinement levels in the experiments of Section~\ref{sec:cases}, which we present below.

\section{Case Studies} \label{sec:cases} We consider the classical multimachine model\cite{5264042} with stochastic fluctuations in the power injections. In particular, we select the noise model to be driven by an OU process, which has been reviewed in~\cite{9380527} to capture short-term power system dynamics. Furthermore, we impose a constant correlation structure among the connected machines.

Let $(\mathcal{V},\mathcal{E})$ denote the set of nodes and edges describing the topology of the power network comprising $n$ buses. Furthermore, we denote the voltage magnitude, voltage phase angle, and angular velocity of the generators, respectively, as $v_i(t)$, $\delta_i(t)$, $\omega_i(t)$ for $1\le i\le n$. The dynamics is governed by the second-order swing equations with noise fluctuations:
\begin{equation} \label{eqn:swing-equations}
    \begin{split}
        \dot{\delta_i}(t)
        &= \omega_i(t) - \omega_R \\
        \frac{2h_i}{\omega_R}\dot{\omega}_i(t) + d_i\omega_i(t)
        &= p_i^m - p_i^e(t) + \eta_i(t),
    \end{split}
\end{equation} where $p^m_i$ represents the (constant) mechanical power input into each generator and $\eta_i(t)$ represents the noise injected at machine $i$. Furthermore, we define the net active power injection:
\begin{equation} \label{eqn:random-dynamical-system}
    \begin{split}
        p_i^e(t)
        &:= \sum_{i=1}^nb_{ij}v_i(t)v_j(t)\sin(\delta_i(t)-\delta_j(t)) \\ 
        &+ \sum_{i=1}^ng_{ij}v_i(t)v_j(t)\cos(\delta_i(t)-\delta_j(t)).
    \end{split}
\end{equation}   

In vector notation, the complete stochastic model is specified as the following:
\begin{equation} \label{eqn:multimachine-model}
\begin{split}
    \dot{\boldsymbol{v}}(t) 
    &= \mathbf{0} \\
    \dot{\boldsymbol{\omega}}(t)
    &= \frac{\omega_R}{2}\boldsymbol{h}^{-1}\odot\bigg[ 
        -(\boldsymbol{\omega}(t) - \omega_R) \odot \boldsymbol{d} + \boldsymbol{p}^m \\
        &- \bigg\{ 
            [\boldsymbol{G}\odot \cos(\Delta) + \boldsymbol{B}\odot \sin(\Delta)]\boldsymbol{v}(t)
        \bigg\} \odot \boldsymbol{v}(t) + \boldsymbol{\eta}(t)] \\
        \dot{\boldsymbol{\delta}}(t) 
    &= (\boldsymbol{\omega}(t) - \omega_R) \\
        \dot{\boldsymbol{\eta}}(t) 
    &= -\theta\boldsymbol{\eta}(t) + \alpha\sqrt{2\theta}\boldsymbol{C}\cdot \frac{d\boldsymbol{W}(t)}{dt} .
\end{split}
\end{equation} 

In the system (\ref{eqn:multimachine-model}), drift and diffusion parameters are taken as $\theta=1$ and $\alpha=0.05$, respectively. $\odot$ denotes Hadamard (elementwise) product. Let $\boldsymbol{v}(t) =[v_1(t), \ldots, v_n(t)]^T$ denote the voltage magnitude of the machines, which are taken as randomly perturbed around a mean operating state that does not have additional temporal dynamics, derived from the constant impedance loads assumption in~\cite{5264042}. $\boldsymbol{\omega}(t) = [\omega_1(t),\ldots,\omega_n(t)]^T$ and $\boldsymbol{\delta}(t) = [\delta_1(t),\ldots,\delta_n(t)]^T$ respectively denote the vector of speeds and angles of the machines. $\omega_R$ denotes the reference bus speed. $\boldsymbol{\eta}(t)$ denotes the OU noise processes that are correlated across the machines according to the correlation matrix $\boldsymbol{R} := \boldsymbol{CC}^T$. Furthermore
\begin{equation}
    {\boldsymbol{h}}^{-1} = 
    \begin{bmatrix}
        1/h_1 \\
        \vdots \\
        1/h_n
    \end{bmatrix}, \quad 
    \boldsymbol{d} =
    \begin{bmatrix}
        d_1 \\
        \vdots\\
        d_n
    \end{bmatrix}, \quad
    \boldsymbol{p}^m = 
    \begin{bmatrix}
        p_1^m\\
        \vdots\\
        p_n^m
    \end{bmatrix},
\end{equation} denote, respectively, the vector of inverse inertia coefficients, constant damping factors, and constant equilibrium total power injections. Finally
\begin{equation}
    \boldsymbol{G} := [g_{ij}], \quad \boldsymbol{B} := [b_{ij}],
\end{equation} are the constant real and imaginary parts of the admittance matrix, respectively. We define
\begin{equation}
    \cos(\Delta) := 
    \begin{bmatrix}
        \cos(\Delta_{11}) & \cos(\Delta_{12}) & \cdots & \cos(\Delta_{1n}) \\
        \cos(\Delta_{21}) &  \cos(\Delta_{22}) & \cdots & \cos(\Delta_{2n}) \\
        \vdots & \cdots & \ddots & \vdots \\
        \cos(\Delta_{n1}) & \cos(\Delta_{n2}) & \cdots & \cos(\Delta_{nn}) ,
    \end{bmatrix}
\end{equation} to be the result of cosine applied elementwise on $\Delta$, and $\sin(\Delta)$ is similarly defined, where
\begin{equation}
    \Delta_{ij}(t) := \delta_i(t) - \delta_j(t)
\end{equation} denotes the pairwise angle difference. Letting $\boldsymbol{z}(t) := [\boldsymbol{v}(t)^T, \boldsymbol{\omega}(t)^T, \boldsymbol{\delta}(t)^T, \boldsymbol{\eta}(t)^T]^T$, the effective dimension of the system (\ref{eqn:multimachine-model}) is $N=4n$. 

\subsection{Density of Line Energy} \label{sec:density-of-one-line} In the study of reliable power systems, stability of a transmission network can be quantified and studied through a closed-form energy function that is dependent on the system states of the dynamic model \cite{5618968}. Subsequently, line failure events are typically modeled as random and rare events that occur in a network during operation. Upon stochastic perturbation, the energy function associated with a particular line may exceed a certain user-specified safety threshold, upon which a relay action is triggered, causing the line to be removed from graph $(\mathcal{V},\mathcal{E})$. A modification of the admittance matrices reflects such removals, and in turn effectively changes the system~(\ref{eqn:multimachine-model}). The post-relay dynamics continues to be monitored until the system states transition and settle in another equilibrium \cite{5618968}.

As a step toward efficient prediction of failure rates using a stochastic power system model, it is of interest to characterize the probability density of line energy functions. Specifically, assuming the baseline model (\ref{eqn:multimachine-model}), let $l\in \mathcal{E}$ denote an edge corresponding to generator bus indices $(i,j)$. We define the quantity of interest as the magnitude of current flowing through line $l$:
\begin{equation} \label{eqn:line-energy-multimachine}
\begin{split}
    u(\boldsymbol{z}(t)) := \frac12b_{ij}^2\big[&(v_i(t))^2 - 2v_i(t)v_j(t)\cos\big(\delta_i(t) - \delta_j(t)\big) \\
    &+ (v_j(t))^2\big] .
\end{split}
\end{equation}

Following the framework derived in Section~\ref{sec:1dropdf-derive}, we can compute the following response variables explicitly:
\begin{equation} \label{eqn:line-energy-drift}
    \mu^u := b_{ij}^2v_i(t)v_j(t)\sin\big(
        \delta_i(t) - \delta_j(t)
    \big)(\omega_i(t) - \omega_j(t)),
\end{equation} with $\mathcal{D}^u \equiv 0$ due to cancellation of the second-order term in  (\ref{eqn:reduced-order-pdf-equation}). The precise derivations of the response variables in (\ref{eqn:line-energy-drift}) are provided in Section~\ref{sec:derive-response-variables} of the Appendix. 
Upon substituting (\ref{eqn:line-energy-drift}) into the reduced-order PDF equation (\ref{eqn:reduced-order-pdf-equation}), we obtain a 1D advection equation:
\begin{equation} \label{eqn:line-energy-ropdf}
    \frac{\partial f_u}{\partial t} + b_{ij}^2\frac{\partial}{\partial U}\big(
        \mathcal{R}(U,t)f_u
    \big) = 0 ,
\end{equation} where the advection coefficient has the following formally exact expression, according to equation (\ref{eqn:reduced-order-pdf-equation}):
\begin{equation} \label{eqn:line-energy-regression-func}
\begin{split}
    \mathcal{R}(U,t) = \mathbb{E}\big[v_iv_j\sin\big(\delta_i-\delta_j\big)(\omega_i-\omega_j) \big| 
     u(t)=U],
\end{split}
\end{equation} which is to be estimated from trajectory data depending on realizations of (\ref{eqn:line-energy-multimachine}) via solving the minimization problem (\ref{eqn:l2-regression}), as outlined in Algorithm~\ref{alg:mainalg}.

\subsection{Joint Dynamics of Line Pairs} \label{sec:joint-line-failures} When a line fails, its voltage load is redistributed to adjacent generators, causing an unexpected spiking of voltages in the local topology that may potentially cause expansive disruptions. For this reason, conditional failure events of adjacent machines are of great practical interest \cite{roth2019kinetic}. We further consider the joint probability of two lines simultaneously exceeding their respective safety thresholds, which incorporates the conditional probabilistic structure.

Let two non-islanding lines $l_1,l_2\in\mathcal{E}$ correspond to machines $(i,j)$ and $(j,k)$ with node $j$ being an intermediate node from node $i$ to node $k$. We follow the generalized formulation in (\ref{eqn:m-dim-ropdf}) and define the vector-valued quantity of interest, which is composed of the energy functions of both lines $l_1, l_2$:
\begin{equation} \label{eqn:joint-line-energy-drift}
{\boldsymbol{\mu}}_{\boldsymbol{u}}=
\begin{bmatrix}
    b_{ij}^2v_iv_j\sin(\delta_i-\delta_j)(\omega_i-\omega_j) \\
    b_{jk}^2v_jv_k\sin(\delta_j-\delta_k)(\omega_j-\omega_k)
    \end{bmatrix}
\end{equation} with no additional diffusion, similar to that of the 1D case; that is, $\mathcal{D}^{\boldsymbol{u}} \equiv \mathbf{0}$. In conjunction with the results of (\ref{eqn:m-dim-ropdf}), the above derivation yields an advection equation in 2D:
\begin{equation} \label{eqn:joint-ropdf-equation}
\begin{split}
    \frac{\partial f_{\boldsymbol{u}}}{\partial t} 
    &+ 
    b_{ij}^2\frac{\partial}{\partial U_1}\big( 
        \mathcal{R}_{1}(\boldsymbol{U},t)f_{\boldsymbol{u}}
    \big)\\
    &+
    b_{jk}^2\frac{\partial}{\partial U_2}\big(
        \mathcal{R}_{2}(\boldsymbol{U},t)f_{\boldsymbol{u}}
    \big) = 0,
\end{split}
\end{equation} where the phase vector $\boldsymbol{U} = [U_1,U_2]$, and the advection coefficients are now bivariate functions, defined as the following: 
\begin{equation} \label{eqn:2d-cond-exp}
\begin{split}
    \mathcal{R}_1(\boldsymbol{U},t) &= \mathbb{E}[v_iv_j\sin(\delta_i-\delta_j)\\&(\omega_i-\omega_j)|u_{1}(t)=U_1,u_2(t)=U_2]
\end{split}
\end{equation}
\begin{equation} \label{eqn:2d-cond-exp-in-y}
\begin{split}
    \mathcal{R}_2(\boldsymbol{U},t) &= \mathbb{E}[v_jv_k\sin(\delta_j-\delta_k)\\&(\omega_j-\omega_k)|u_{1}(t)=U_1,u_2(t)=U_2] .
\end{split}
\end{equation}

The conditional expectations 
\eqref{eqn:line-energy-regression-func}, \eqref{eqn:2d-cond-exp}, and \eqref{eqn:2d-cond-exp-in-y}
will be estimated by regression, and the resulting advection-diffusion equations 
\eqref{eqn:line-energy-ropdf} in 1D and \eqref{eqn:joint-ropdf-equation} in 2D will be computed numerically, thus enabling propagation of uncertainty.

\section{Numerical Experiments}
\subsection{Experimental Setup}\label{sec:1dropdf} The experiments were conducted using three test cases: the WSCC 9-bus system, IEEE 30-bus system, and IEEE 57-bus system. To mimic the instability caused by sudden failing of a transmission line during normal operation, we first initialize the power system states with the connections intact. Namely, we disregard the noise model in (\ref{eqn:multimachine-model}) and solve the deterministic optimal power flow (OPF) problem for an equilibrium voltage magnitude $\boldsymbol{v}^* = [v_1^*, \ldots, v_n^*]^T$ and machine angles $\boldsymbol{\delta} = [\delta_1^*, \ldots, \delta_n^*]^T$ using Matpower 7.1~\cite{5491276}. After obtaining equilibrium voltages and angles, we add a small random perturbation in the voltage magnitudes and treat the angles along with prespecified initial speeds at the reference level  $\omega_R=1$, as initial conditions for the random dynamical system (\ref{eqn:random-dynamical-system}). Specifically, the initial conditions for system (\ref{eqn:multimachine-model}) are taken to be
\begin{equation} \label{eqn:sde-initial-cond}
\begin{split}
    \boldsymbol{v}(0) &=
    \abs{\mathcal{N}(\boldsymbol{v}^*, 0.01\cdot\sigma_{{\boldsymbol{v}^*}}^2\boldsymbol{I}_n)} \\
    \boldsymbol{\omega}(0) &= \omega_R\cdot\mathbf{1} \\
    \boldsymbol{\delta}(0) &= \boldsymbol{\delta}^* \\
    \boldsymbol{\eta}(0) &\sim \mathcal{N}(
        \mathbf{0}, 
        \alpha^2\boldsymbol{R} ,
    )
\end{split}
\end{equation} where the initial random voltage magnitudes are generated from a folded Gaussian distribution with mean at the equilibrium levels $\boldsymbol{v}^*$, and perturbed around the OPF equilibrium with strength $0.1\cdot \sigma_{\boldsymbol{v}^*}$. In particular, $\sigma_{\boldsymbol{v}^*}$ is the standard deviation of equilibrium voltages across machines. Following \cite{maltba2022learning}, we take $\alpha = 0.05$, and the correlation matrix $\boldsymbol{R}$ is chosen for simplicity to be constant with moderate correlation levels, varying by application. We use $R_{ij} = R_{ji} = 0.44$ for the WSCC 9-bus case, for all $i, j \in \mathcal{V}, i\neq j$, and $R_{ij}=R_{ji} = 0.36$ in the IEEE 30-bus and 57-bus test cases. 

With the noise model set, we first perform a ``burn-in'' period lasting time $T=50.0$ and observe that the system (\ref{eqn:multimachine-model}) settles in an approximate (stochastic) equilibrium. We then simulate a single line failure by manually setting entries of the admittance matrices $\boldsymbol{B}, \boldsymbol{G}$ to zero. The samples at the final time of the burn-in period are considered to be at equilibrium and again taken as initial conditions to be re-solved after line removal, for an additional time period $T = 10$. To investigate the energy changes caused to adjacent lines in the local topology, we compute and record the trajectories of energy function $\boldsymbol{u}$, defined in (\ref{eqn:line-energy-multimachine}).

In our experiments we specifically choose non-islanding edges in the network and observe the adjacent lines connected to a shared generator. Concretely, lines 8--9 were removed for the WSCC 9-bus case, and energy of lines 4--9, 7--8 were subsequently recorded. Lines 6--8 were removed from IEEE case 30, and energy data was collected for lines 6--7, 6--9. For IEEE case 57, lines 36--37 were removed, and data was collected at lines 35--36 and 36--40. The system (\ref{eqn:multimachine-model}) was integrated by using a Milstein scheme \cite{Kloeden1992} with a uniform time step size of $\Delta t = 10^{-2}$. In particular, we investigate the method's ability to predict probabilities of random events defined with respect to the PDF. The reported 1D and 2D experiments were compared with a benchmark computed from a Gaussian kernel density estimator  from $m_{\text{KDE}} = 10,000$ Monte Carlo trials, using Silverman's rule for bandwidth selection. When reporting probabilities, we use instead the empirical cumulative distribution function, which represents a probability estimate from Monte Carlo sampling, a common approach in conventional studies. Across all test cases, the Monte Carlo sample size used for computing the initial condition, as well as the coefficient estimation (\ref{eqn:l2-regression}), is taken to be $m_{\text{R}}=5,000$, for reported experiments of Section~\ref{sec:1dproblem} and Section~\ref{sec:2dproblem}. As a quantitative measure, we define and henceforth refer to error as the space-time $L^1$ distance from the benchmark
\begin{equation} \label{eqn:l1-error-measure}
    e(\widehat{f}, f) \equiv \int_{0}^T\int_{\Omega}\abs{\widehat{f}(\boldsymbol{U},t) - f(\boldsymbol{U},t)}d\boldsymbol{U}dt,
\end{equation} where, for notation consistency, $\widehat{f}$ is understood to be the result of the reduced-order PDF method (\ref{eqn:m-dim-ropdf}) and $f$ is the KDE benchmark. The computational domain $\Omega\subset\mathbb{R}^{N_{\text{R}}}$ is determined from ranges of simulated values of $\boldsymbol{u}(t)$, with a padding on the boundaries of $\pm 0.5$-$1$ standard deviations. With the numerical domain set, we approximate the vanishing boundary conditions at infinities, as required by (\ref{eqn:general-fpe}), with a Dirichlet boundary condition. Furthermore, to compare and evaluate the accuracy of probability estimations of a predefined event $E\subset\Omega$, we also report the probabilities $\widehat{\mathbb{P}}(E), \mathbb{P}(E)$, where by construction $\widehat{\mathbb{P}}(E) = \int_E\widehat{f}$ is a prediction output from the reduced-order PDF and $\mathbb{P}$ is the benchmark result computed from empirical CDF. In Section~\ref{sec:sample-complexity} and Section~\ref{sec:mutual-information} we examine empirically the sample scaling and correlation structure, respectively. 

\begin{figure*}
    \centering
    \includegraphics[width=7cm]{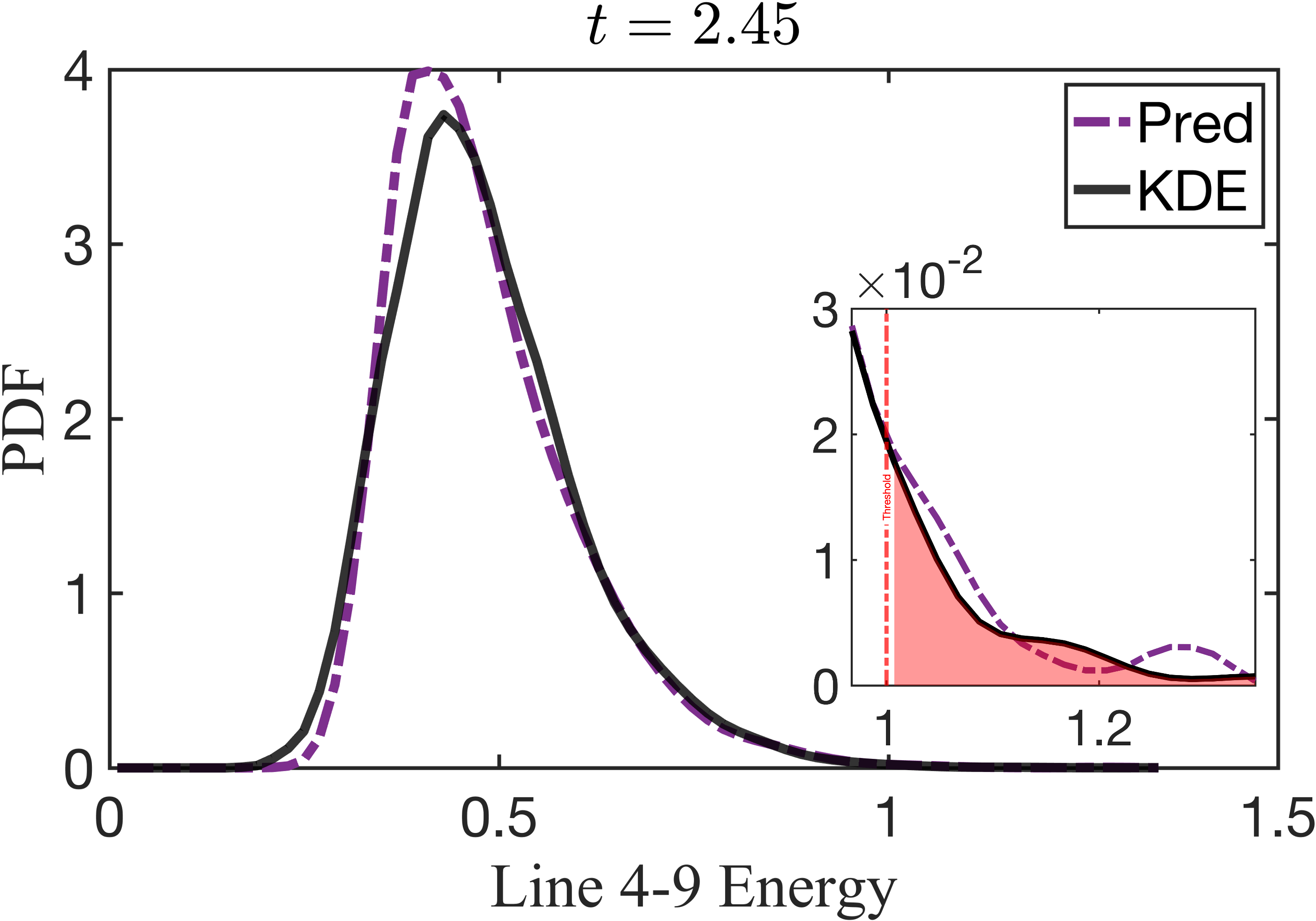}
    \includegraphics[width=7cm]{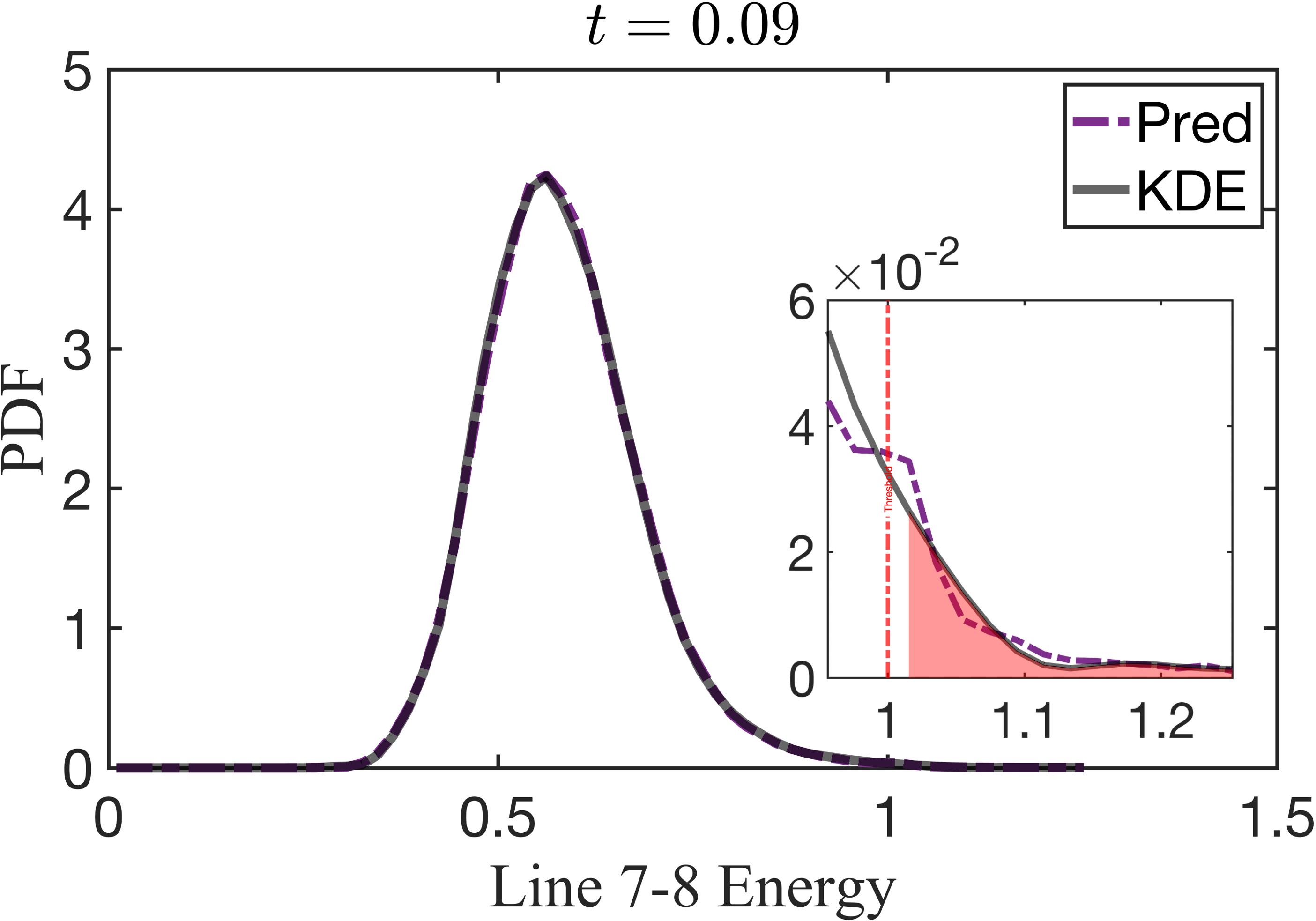}
    \includegraphics[width=6.7cm]{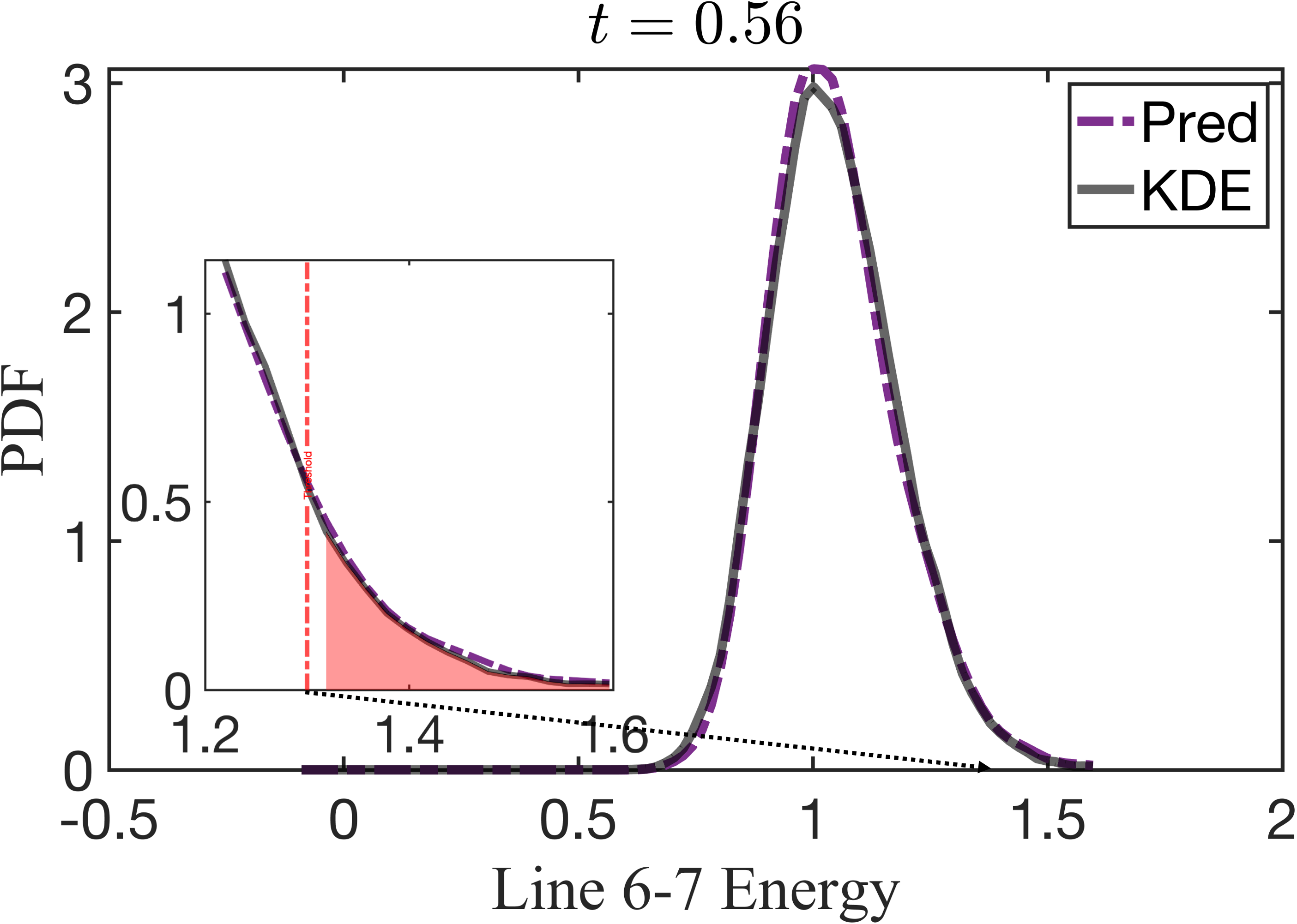}
    \includegraphics[width=7cm]{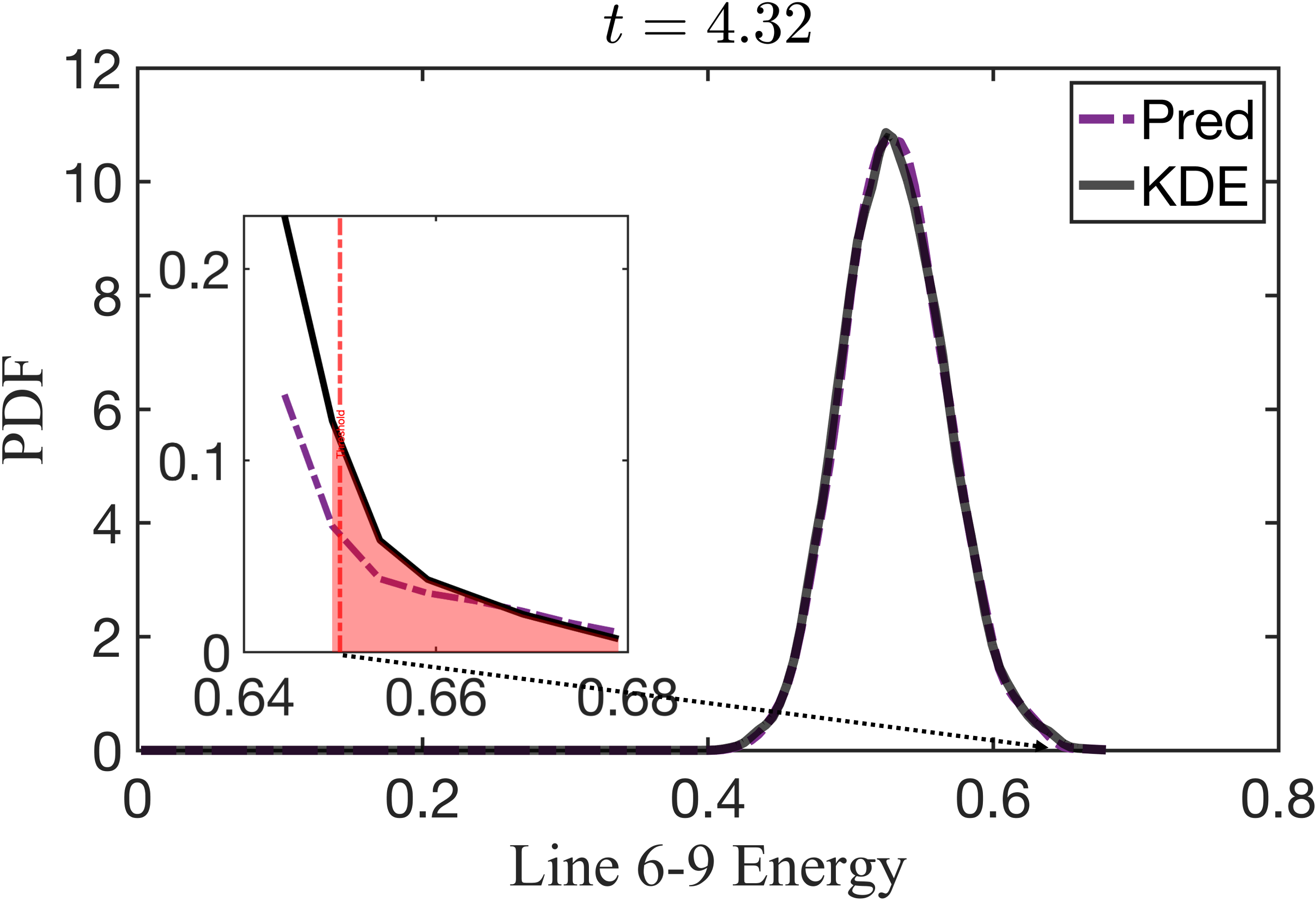}
    \includegraphics[width=7cm]{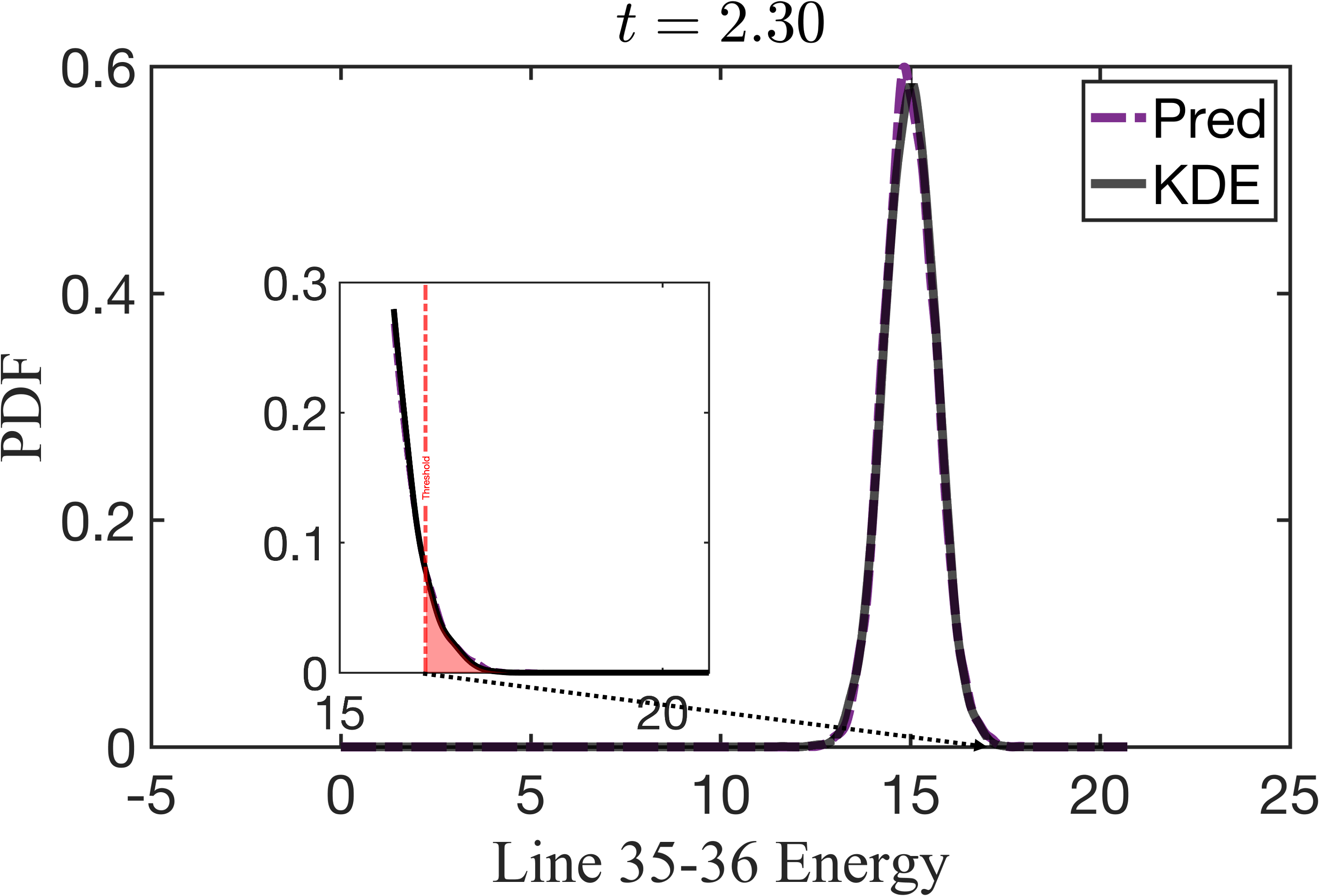}
    \includegraphics[width=7cm]{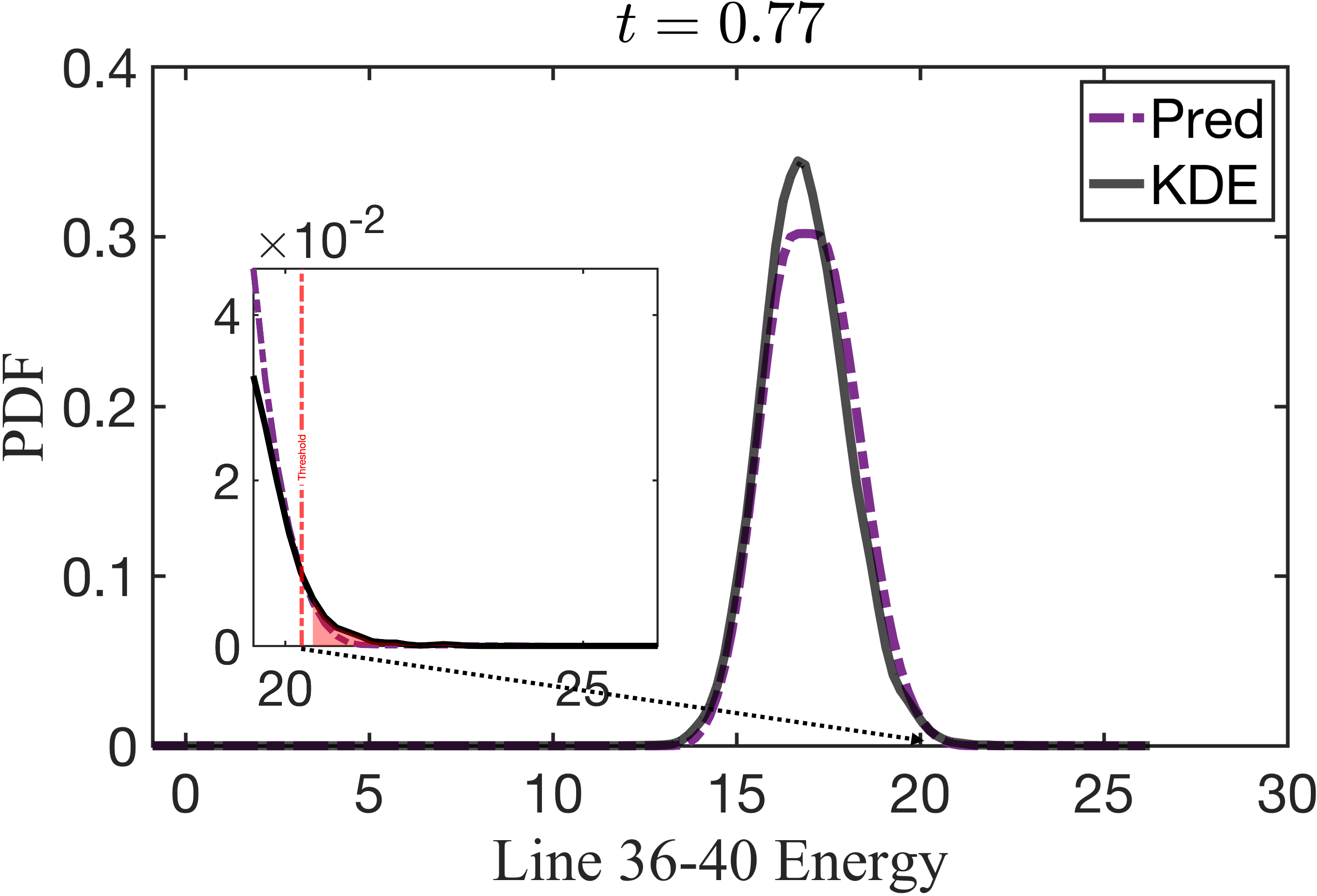}
    \caption{(Top to bottom) Predicted PDF of line energies for WSCC case 9, IEEE case 30, and IEEE case 57, plotted at respective peak times, with exceedance thresholds zoomed in and marked in red. The 1D reduced-order marginal PDF equations (\ref{eqn:coord-projection-qoi}) were solved with $m_{\text{R}} = 5,000$ samples, compared with a $m_{\text{KDE}} = 10,000$ KDE benchmark. }
    \label{fig:ropdf1d}
\end{figure*}

\subsection{Marginal Probability Densities} \label{sec:1dproblem}

To validate our proposed method, we first solve the 1D reduced-order PDF equations (\ref{eqn:joint-ropdf-equation}). The reduced-order PDF equations marginal in selected line energies, as outlined in Algorithm~\ref{alg:mainalg}, were solved by using a standard finite volume Lax--Wendroff method with a Monte Carlo flux limiter \cite{leveque_2002}. To resolve the unknown terms in (\ref{eqn:line-energy-regression-func}), we used Gaussian local linear regression with 10-fold cross validations. The choice of regression method is dependent on the underlying system dynamics at hand and, naturally, requires a degree of manual tuning. The analysis of effects of regression methods is an open problem and can be studied via systematic model validation~\cite{76a2be57-3fd6-37ca-8baa-f26d697c3f93}. 

\begin{figure}
    \centering
    \includegraphics[width=8cm]{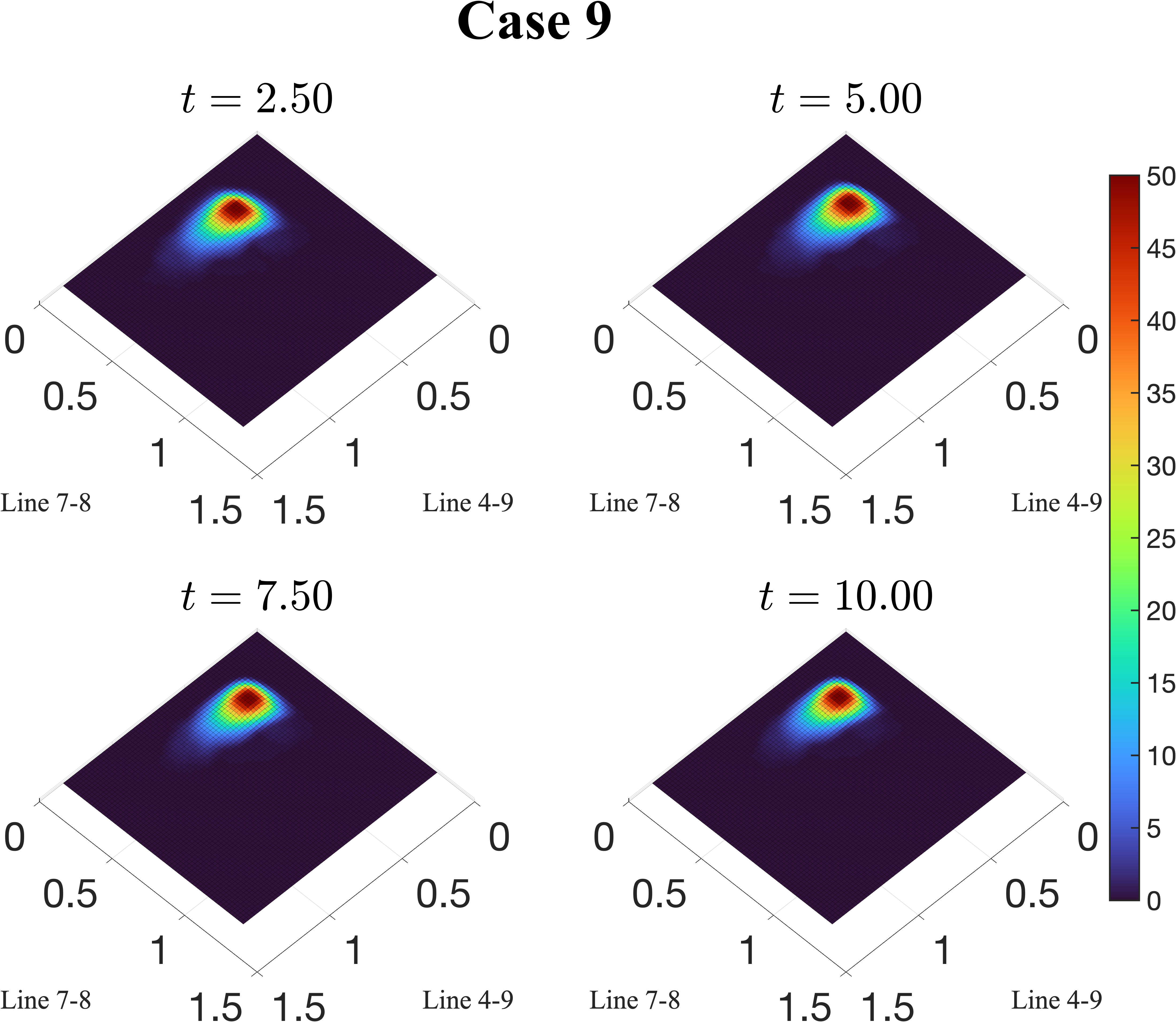}
    \includegraphics[width=8cm]{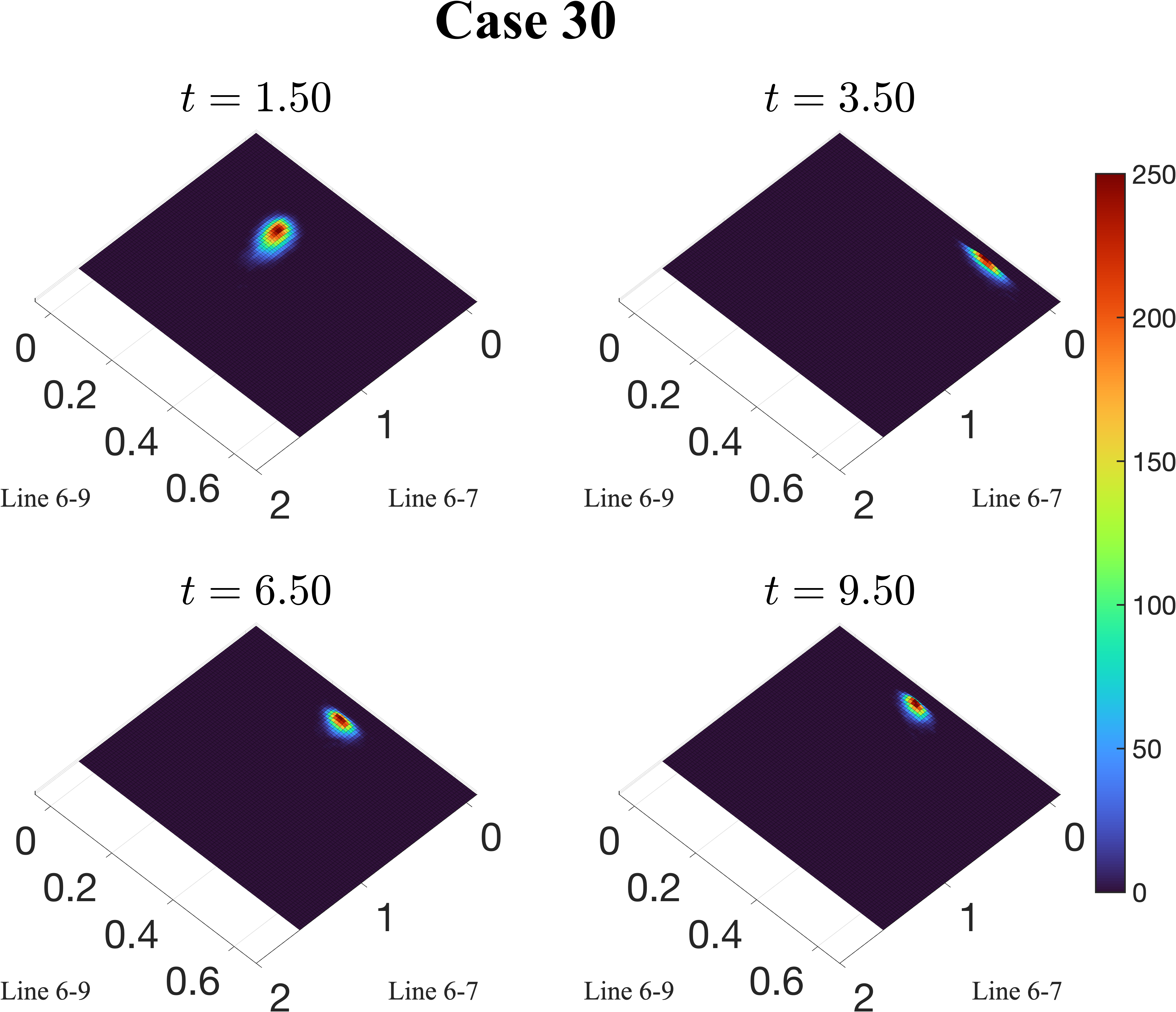}
    \includegraphics[width=8cm]{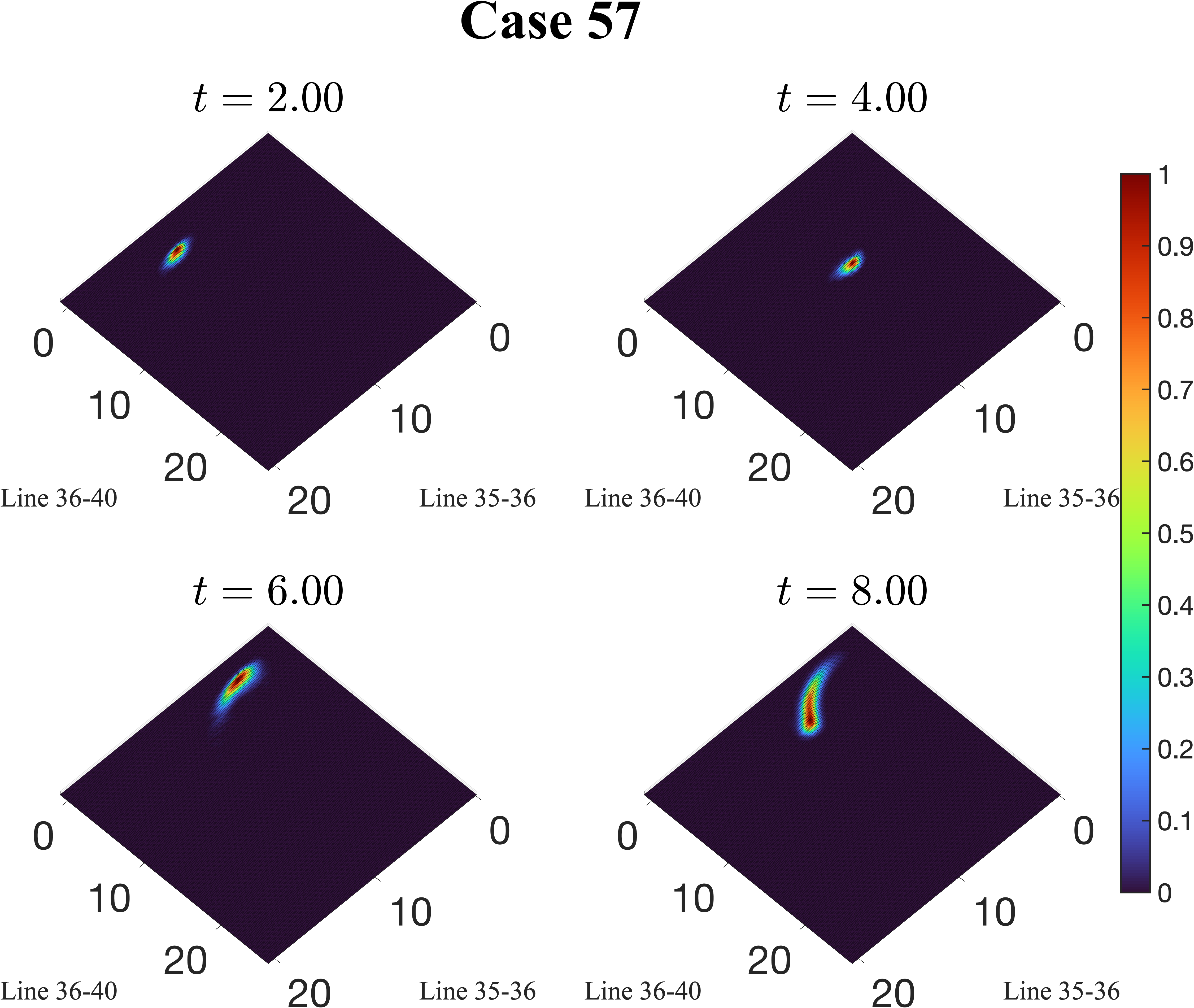}
    \caption{Predicted reduced-order joint PDFs (\ref{eqn:joint-ropdf-equation}) with $m_{\text{R}} = 5,000$ samples, at selected PDE solution times. Particularly, IEEE Case 30 and 57 show non-trivial evolution and correlation structure over time. }
    \label{fig:tailprob2d}
\end{figure}

We show in Figure~\ref{fig:ropdf1d} the predicted PDF computed from the reduced-order equation by numerical integration, plotted at ``peak time'' where the energy level (averaged over all $m_{\text{R}}=5,000$ realizations) for each line reaches its maximum during simulation time. The predicted PDF is compared in the same plot with KDE using $m_{\text{KDE}}=10,000$ samples. As a potential use case, we focus  on the tail region defined by the exceedance of long-term line rating (specified in \texttt{Matpower} as field \texttt{rateA}). Specifically, we consider the following time-parameterized event:
\begin{equation} \label{eqn:individual-event}
    E_l(t) \equiv \{u_l(t) > u_{l,\text{max}}\},
\end{equation} where $u_{l,\text{max}}$ is understood as the individual line rating for line $l$. The probability of event $E_l$ can be readily computed through numerical integration of the predicted PDF $\widehat{f}$, which we report in Table~\ref{tbl:tail-probabilities-reported}. The predicted tail event probabilities were compared with those provided by an empirical CDF evaluation, which corresponds to the conventional Monte Carlo integral computed directly using samples of $u_l(t)$. In the investigated lines, lines 4--9, 7--8 of WSCC case 9 have a prespecified rating of 1 p.u. Lines 6--7, 6--9 in the IEEE case 30 have, respectively, ratings of 1.3 and 0.65 p.u.. For the IEEE case 57, the line ratings were not provided by \texttt{Matpower} but instead were estimated from solving the OPF with the Julia \texttt{PowerModels.jl} library using the \texttt{\_calc\_thermal\_limits} function. The obtained ratings for lines 35--36, 36--40 were 16.33 and 20.27 p.u., respectively. From both Figure \ref{fig:ropdf1d} and Table \ref{tbl:tail-probabilities-reported} we observe good agreement between our method and KDE estimation (the latter using double the number of samples as the former).

\subsection{Joint Failure Probabilities} \label{sec:2dproblem} Following the discussions in Section~\ref{sec:joint-line-failures}, we solved the 2D reduced-order PDF equations joint in lines $l_1, l_2$ using a corner transport upwind  method with second-order corrections and a van Leer flux limiter~\cite{leveque_2002}. With the same computational domain setup as the 1D cases in Section~\ref{sec:1dproblem} to approximate vanishing boundary, we applied both ordinary linear regression and locally weighted linear regression (Lowess) at each step of PDE integration to obtain the regression functions (\ref{eqn:2d-cond-exp}) from observations of $\boldsymbol{u}(t)$, with the latter being much more computationally demanding. The reduced-order equations were solved by using $m_{\text{R}} = 5,000$ Monte Carlo samples and were compared with a benchmark computed from bivariate Gaussian KDE with $m_{\text{KDE}}=10,000$ Monte Carlo samples. Figure~\ref{fig:tailprob2d} visualizes selected temporal snapshots of the predicted PDFs. In particular, we observe that the joint PDF of WSCC case 9 is relatively stationary over time, whereas that of IEEE case 30 and case 57 show a high degree of variation.

\begin{figure*}[!ht]
    \centering
    \begin{tabular}[b]{r rrr}
        \toprule              
        & \mc3c{Case 9 ($t=2.45$)}$1\times 10^{-3}$ \\
            \cmidrule(r){2-4} 
        & Lines               
        & Predicted
        & Empirical        \\
        \midrule
        & $(4,9)$ 
            & $1.671$
            & $1.400$ \\
        & $(7,8)$ 
            & $1.496$ 
            & $1.400$ \\
        & Joint 
            & {\color{gray}
              $0.658$
            } / $0.659$
            & $0.458$ \\ 
    \end{tabular} \qquad
    \includegraphics[width=7cm]{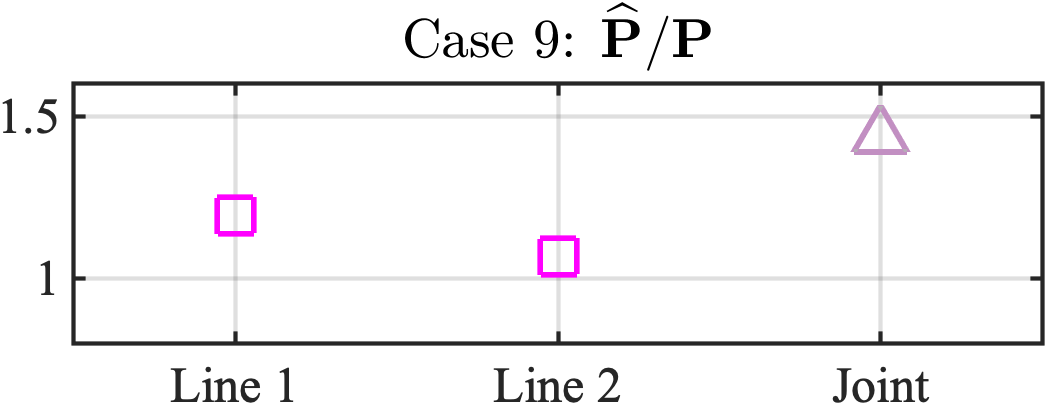}
    \begin{tabular}[b]{r rrr}
        \toprule              
        & \mc3c{Case 30 $(t=0.56)$}$1\times 10^{-2}$ \\
            \cmidrule(r){2-4} 
        & Lines               
        & Predicted
        & Empirical        \\
        \midrule
        & $(6,7)$
            & $3.625$ 
            & $3.270$\\
        & $(6,9)$
            & $0.058$
            & $0.060$\\
        & Joint 
            & {\color{gray}
              $5.580$
            } / $2.110$ 
            & $2.112$\\ 
    \end{tabular} \qquad
    \includegraphics[width=7cm]{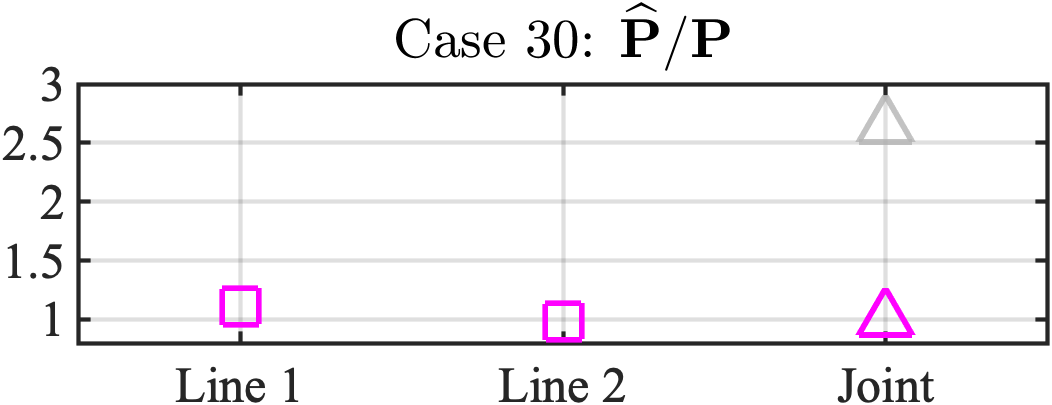}
    \begin{tabular}[b]{r rrr}
        \toprule              
        & \mc3c{Case 57 ($t=2.30$)}$1\times 10^{-2}$ \\
            \cmidrule(r){2-4} 
        & Lines               
        & Predicted
        & Empirical        \\
        \midrule
        & $(35,36)$ 
            & $2.968$ 
            & $2.760$    \\
        & $(36,40)$ 
            & $2.424$ 
            & $1.970$    \\
        & Joint 
            & {\color{gray}
              $2.056$
            } / $0.918$ 
            & $1.029$    \\ 
                \bottomrule
    \end{tabular} \qquad
    \includegraphics[width=7cm]{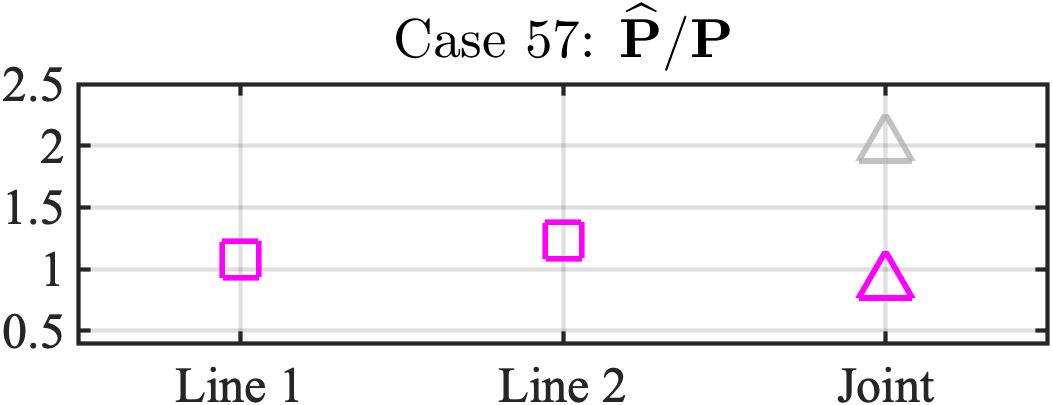}
    \captionsetup{labelformat=andtable}
    \caption{
        (Left) All test cases reported at energy peak time, tail probabilities for individual exceedance (\ref{eqn:individual-event}), and joint exceedance (\ref{eqn:tail2d-event}). The values given by reduced-order PDF equations are computed with $m_{\text{R}} = 5,000$ samples and compared with those obtained from $m_{\text{KDE}} = 10,000$ KDE benchmark. The tail probabilities estimated by using independence assumption (\ref{eqn:independence-assumption}) are marked in grey.
        (Right) Ratios between predicted probability $\widehat{\mathbb{P}}$ and benchmark probability $\mathbb{P}$, as described in Section~\ref{sec:1dropdf}. The grey markers indicate the computation when the line failures are assumed to be independent, which is significantly farther away from 1 (i.e. away from matching the empirical observation of the frequency of two simultaneous events).
    }
    \label{tbl:tail-probabilities-reported}
\end{figure*}

To evaluate the joint probabilistic profile through a tail event, we let $U_{l_1,\text{max}}, U_{l_2,\text{max}}$ denote the respective constant line ratings, and we consider the region where either of the two lines fail, i.e., a pure series failure criterion:
\begin{equation} \label{eqn:tail2d-event}
    E_{l_1,l_2}(t) \equiv \{u_{l_1}(t)>U_{l_1,\text{max}}\cup u_{l_2}(t)>U_{l_2,\text{max}}, \}
\end{equation} which can be evaluated from the predicted joint PDF $\widehat{f}_{l_1l_2}$ from equation (\ref{eqn:joint-ropdf-equation}), by integrating the following numerically:
\begin{equation*} 
    \begin{split}
        \widehat{\mathbb{P}}(E_{l_1,l_2}) 
            &= 1 - \widehat{\mathbb{P}}(\{u_{l_1}\le U_{l_1,\text{max}}\cap u_{l_2}\le U_{l_2,\text{max}}\}) \\
            &= 1-\int_{-\infty}^{U_{l_1,\text{max}}}\int_{-\infty}^{U_{l_2,\text{max}}}\widehat{f}_{l_1l_2}(U_1,U_2,t)dU_1dU_2 .
    \end{split}
\end{equation*} The predicted tail probabilities of events (\ref{eqn:tail2d-event}) for all test cases were reported in Table~\ref{tbl:tail-probabilities-reported}, in comparison with values computed from numerically integrating the benchmark KDE. Similarly to the marginal probabilities, we record the probabilities at the peak time at the larger of the two line energies. 

We make a case for the consideration of joint events, rather than assuming the independence of failures. We highlight the deviating results (compared with the benchmark) in Table~\ref{tbl:tail-probabilities-reported} computed from a product of independent marginal probabilities. Namely, the assumption:
\begin{equation} \label{eqn:independence-assumption}
    \begin{split}
        \widehat{\mathbb{P}}(\{u_{l_1}\le U_{l_1,\text{max}}&\cap u_{l_2}\le U_{l_2,\text{max}}\}) = \\
        \widehat{\mathbb{P}}(\{u_{l_1}\le &U_{l_1,\text{max}}\})\cdot \widehat{\mathbb{P}}(\{u_{l_2}\le U_{l_2,\text{max}}\})
    \end{split}
\end{equation} does not hold in general and yields inaccurate representations of failure probabilities, particularly for the cases 30 and 57. This is relevant since the approach used in \cite{roth2019kinetic} would be unable to treat the occurrence of two failure events as dependent, and thus the approach here, while more expensive, presents a significant modeling advantage.

\subsection{Empirical Sample Complexity} \label{sec:sample-complexity} 
\begin{figure}
    \centering
    \includegraphics[width=8cm]{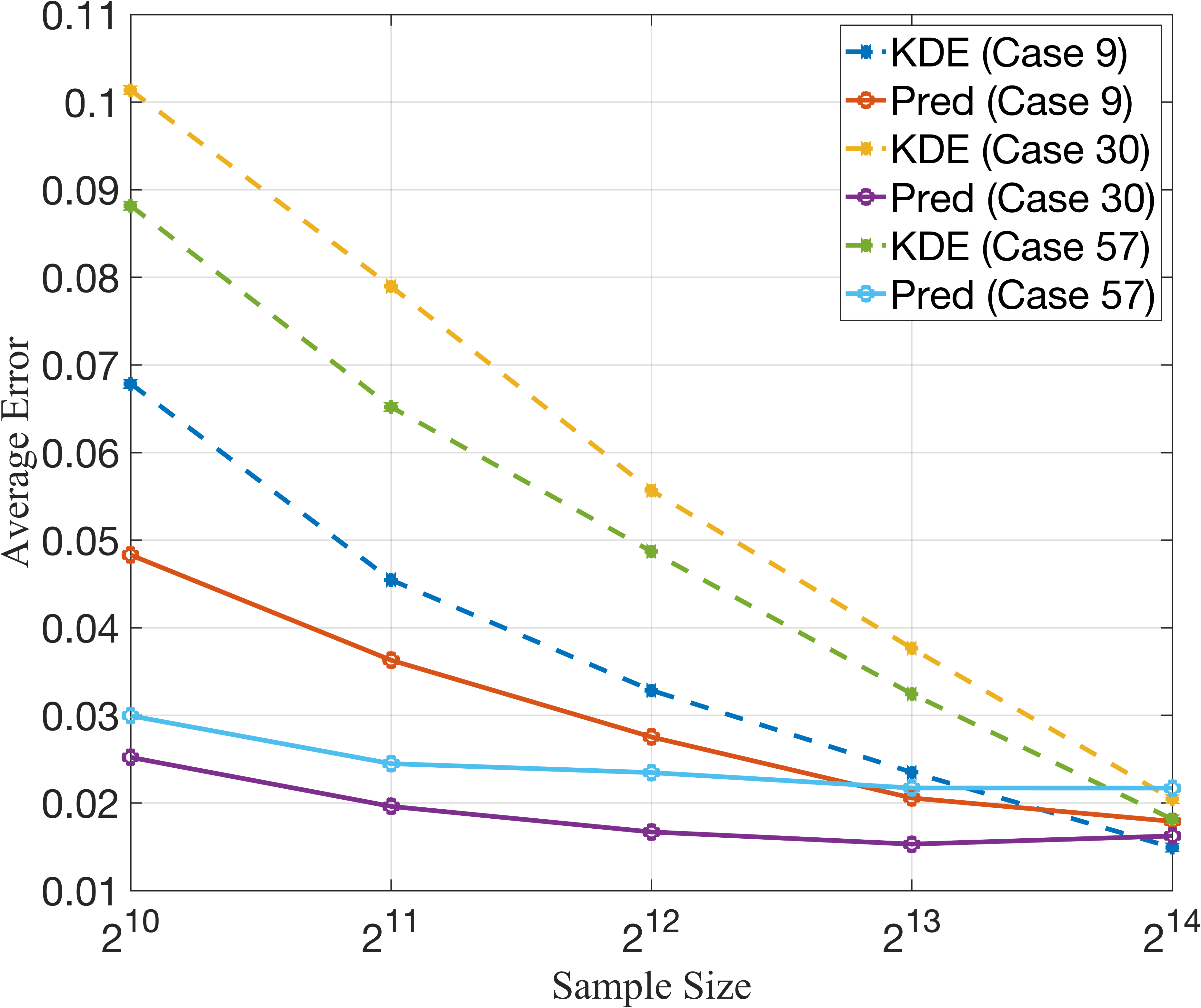}
    \includegraphics[width=8cm]{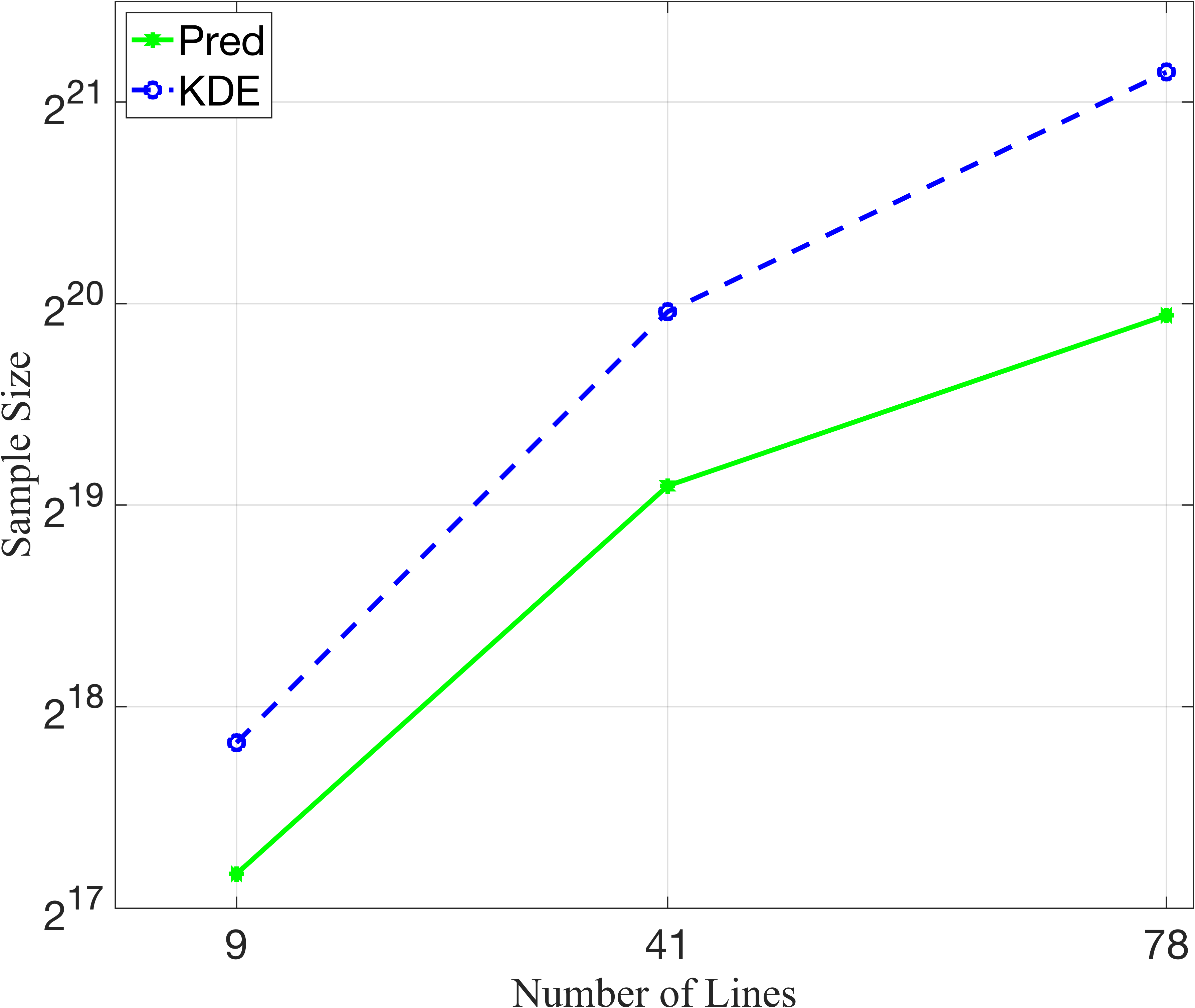}
    \caption{(Top) Convergence of $L^1$-error from benchmark (\ref{eqn:l1-error-measure}), averaged by the number of lines for each case. (Bottom) Aggregate number of samples required for KDE or reduced-order PDF method to reach within $\gamma=1\%$ $L^1$-error from the benchmark.}
    \label{fig:sample-complexity}
\end{figure}

For each test case, we investigate the reduced-order PDF method's practical scalability through parallelizing the computation over all line energies and counting the total number of samples needed for sufficient accuracy. In precise terms, we define the sample complexity measure for each line $l$:
\begin{equation} \label{eqn:total-sample-count}
    m_l^*(\gamma) := 
    \min\{m: e(\widehat{f}_{l}(\cdot; m), f_{l,\text{bench}}) < \gamma, \}
\end{equation} where we let $\widehat{f}_l(\cdot; m)$ denote the reduced-order PDF solution computed with $m$ Monte Carlo samples, to emphasize consideration of sample size. $f_{l,\text{bench}}$ is a benchmark solution computed from a high-resolution KDE using $2^{15} = 327,68$ samples. $m_l^*$ thus quantifies the minimum number of samples required for the predicted PDF of line $l$ to converge within $0<\gamma<1$ in $L^1$-norm (\ref{eqn:l1-error-measure}) of the benchmark KDE. Proceeding, we consider the aggregate sample complexity over all possible lines:
\begin{equation} \label{eqn:total-sample-complexity}
    m^*(\gamma) := \sum_{l=1}^{\abs{\mathcal{E}}}m_l^*(\gamma),
\end{equation} which is a case-dependent quantity due to the increasing number of connections in the power system. Without introducing line tripping, we simulate the system (\ref{eqn:multimachine-model}) in stochastic equilibrium for $T=2.0$ with a refined time step size of $\Delta t = 5\times 10^{-3}$ and with the same noise parameters as in Section~\ref{sec:1dropdf}. Then, sample sizes are increased from $2^{12} = 2,048$ to $2^{14} = 163,84$ in the computations of reduced-order PDF equations (using linear regression for closure estimation (\ref{eqn:line-energy-regression-func})), described in equation (\ref{eqn:line-energy-ropdf}). As a comparison to the reduced-order PDF approach, a low-fidelity (in contrast to the benchmark) Gaussian KDE computed using the same number of samples was recorded for each example. The sequence of increasing sample sizes yield reduced-order PDF solutions and low-fidelity KDE of improved cumulative $L^1$-accuracy in the sense of (\ref{eqn:l1-error-measure}), when compared to the benchmark KDE. Finally, specifying the error threshold $\gamma$, we count and aggregate the number of samples required for each line, in order to obtain a final estimate of total sample complexity (\ref{eqn:total-sample-complexity}). This procedure for investigating sample scalability is summarized in Figure~\ref{fig:sample-complexity}, where we observe a reduction of sample counts for an accuracy requirement of $1\%$. The dependence of sample complexity (\ref{eqn:total-sample-complexity}) on the number of lines $\abs{\mathcal{E}}$ for the reduced-order PDF method has an estimated scaling of $O(\abs{\mathcal{E}}^{0.89})$, compared with $O(\abs{\mathcal{E}}^{1.07})$ for Gaussian KDE. While further sample complexity reduction has been noted in, for instance \cite{maltba2022learning}, where regression functions (\ref{eqn:l2-regression}) admit partial separation (i.e., explicit dependence on QoI $\boldsymbol{u}(t)$ can be factored out due to conditioning and need not be estimated by regression) and thus variance reduction, the reduction in our case is particularly noteworthy because the response variable (\ref{eqn:line-energy-ropdf}) is not an elementary expression of the line energy (\ref{eqn:line-energy-multimachine}).
In any case, the performance improvement of our method relative to direct simulation is significant, as we see from Figure \ref{fig:sample-complexity} that the accuracy of our method with about $2^{10}$ samples can be achieved only at about $2^{14}$ samples with direct Monte Carlo. Since KDE does not naturally impose any time-dependent structure, the empirical result suggests the importance and plausibility of investigating scalable reduced-order models of time-dependent systems.

\subsection{Mutual Information} \label{sec:mutual-information} 
\begin{figure}
    \centering
    \includegraphics[width=8.5cm]{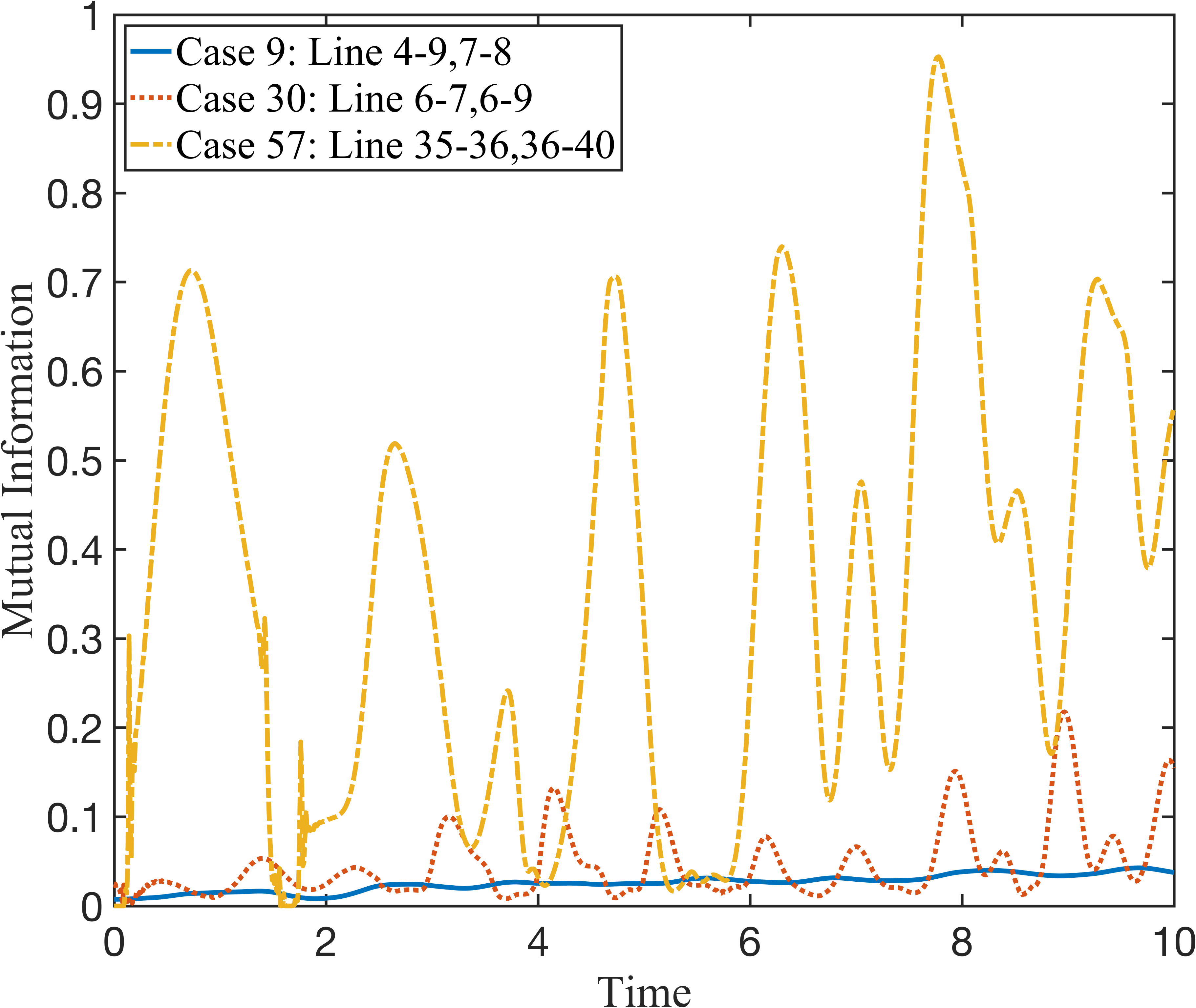}
    \caption{Estimated mutual information (\ref{eqn:mutual-info}) of WSCC case 9 and IEEE cases 30 and 57 over PDE simulation time up to $T=10$. }
    \label{fig:mutual-info}
\end{figure}

Information transmission is studied in Bayesian reliability analysis to identify fragile components of a power network under system perturbations \cite{ZHONG2021118867,7232592}. As an alternative quantification of the correlation structure for the measured lines under load uncertainty, we consider a discrepancy measure between the predicted 2D joint PDF, and the product of marginal PDFs, which represents the density obtained from the independence assumption~(\ref{eqn:independence-assumption}). The Kullback--Leibler (KL) divergence, denoted as $D_{\text{KL}}$, is used to assess the strength of interdependency between pairs of random variables, also known as their mutual information. Total correlation among a larger subset of random states can also be conveniently generalized in graphical models that incorporate the network topology directly; see, for instance~\cite{Studeny1998}. 
For our application, we define in 2D the approximate KL divergence for the energy distributions of lines $l_1,l_2\in\mathcal{E}$:
\begin{equation} \label{eqn:mutual-info}
\begin{split}
    D_{\text{KL}}(\widehat{f}_{l_1l_2}\lvert\rvert \widehat{f}_{l_1}\otimes\widehat{f}_{l_2}; t) := \\ 
    \int\int \widehat{f}_{l_1l_2}(U_1,U_2,t)\log\bigg(
        &\frac{\widehat{f}_{l_1l_2}(U_1,U_2,t)}{\widehat{f}_{l_1}(U_1,t)\widehat{f}_{l_2}(U_2,t)}
    \bigg)dU_1dU_2 ,
\end{split}
\end{equation} where the density $\widehat{f}_{l_1}\otimes\widehat{f}_{l_2}$ denotes the product probability measure from 1D reduced-order PDFs (\ref{eqn:line-energy-drift}). Figure~\ref{fig:mutual-info} shows the plotted curves for all cases predicted using $m_{\text{R}}=5,000$ observations. In particular, Figure~\ref{fig:mutual-info} suggests that the assessed lines in WSCC case 9 and IEEE case 30 are on average only weakly correlated; thus, estimating marginal densities separately may provide a relatively faithful approximation (though at the peak time such discrepancy did manifest more prominently for IEEE case 30, as shown in Table \ref{tbl:tail-probabilities-reported}). The correlation is non-negligible for IEEE case 57, and a product measure approximation does not yield accurate results, as also seen in Table~\ref{tbl:tail-probabilities-reported}.

\section{Conclusions and Future Work} \label{sec:conclude} In this work we investigated the reduced-order PDF method to reformulate the propagation of uncertainty in a power system model as a low-dimensional numerical PDE problem with closure terms that can be modeled empirically by regression. We proposed a general framework that can flexibly incorporate correlated noise models. Furthermore, we provided an implementation and demonstrated the method's ability for describing density of arbitrary QoIs with a considerable reduction of high-dimensional MC simulations. The method is shown to accurately propagate failure probabilities. Most importantly, we demonstrate nontrivial correlation structure particularly for IEEE test cases of increasing sizes, measured through the mutual information (\ref{eqn:mutual-info}), validating the need for and advantage in considering joint densities. Although we have discussed the probabilistic profile of only two lines, the equation for several lines $(N_{\text{R}}\ge 3)$ can be analogously derived. For future work, such a framework holds promise in the prediction of cascading failure events in more complex energy-based power system models involving line tripping dynamics, such as \cite{5618968}. 

\section*{Acknowledgment} 
This material was based upon work supported by the U.S. Department of Energy, Office of Science,
Office of Advanced Scientific Computing Research (ASCR) under
Contract DE-AC02-06CH11347.
\bibliographystyle{abbrv}
\bibliography{ref}
\vspace{12pt}


\section*{Appendix}
\subsection{Derivations of Reduced-Order PDF Coefficients} \label{sec:derive-response-variables}

To make the form of the reduced-order PDF equations considered in Section~\ref{sec:density-of-one-line} and Section~\ref{sec:joint-line-failures} more explicit, we provide a concise computation of the advection and diffusion coefficients that appear in (\ref{eqn:line-energy-drift}), following the formulae introduced in (\ref{eqn:reduced-order-drift}) and (\ref{eqn:reduced-order-diffusion}). Because of the elementary extension to two transmission lines, derivations of the associated coefficients in the joint equation (\ref{eqn:joint-ropdf-equation}) follow exactly.

 We consider line $l\in\mathcal{E}$ and generator indices $(i,j)$ fixed. Furthermore, we recall the energy function defined in (\ref{eqn:line-energy-multimachine}),
 \begin{equation} \label{eqn:recall-energy-function}
    \begin{split}
        u(\boldsymbol{z}(t)) := \frac12b_{ij}^2\big[&(v_i(t))^2 - 2v_i(t)v_j(t)\cos\big(\delta_i(t) - \delta_j(t)\big) \\
        &+ (v_j(t))^2\big] .
    \end{split}
 \end{equation} which is a function of the states evolving according to the noise-perturbed system defined in (\ref{eqn:multimachine-model}). Effectively, only a small subset of variables in $\boldsymbol{z}$ contributes to line $l$'s energy; that is,  $u(\boldsymbol{z}) = u(v_i, v_j, \delta_i, \delta_j)$. As a result, the entries in the gradient, $\nabla_{\boldsymbol{z}}u$, and Hessian matrix, $H_{\boldsymbol{z}}u$, of the line energy vanish except the following contributions of voltage:
\begin{equation} \label{line-energy-derivatives-voltage}
    \begin{split}
        \frac{\partial u}{\partial v_i} &= -\frac{\partial u}{\partial v_j} = b_{ij}^2(v_i - v_j\cos\big(\delta_i - \delta_j)\big) \\
        \frac{\partial^2u}{\partial v_i^2} &= \frac{\partial^2u}{\partial v_j^2} = b_{ij}^2 \\
        \frac{\partial^2 u}{\partial v_i\partial v_j} &= -b_{ij}^2\cos(\delta_i - \delta_j)
    \end{split}
\end{equation} 

And similarly for generator angles:
\begin{equation} \label{line-energy-derivatives-angles}
    \begin{split}
        \frac{\partial u}{\partial\delta_i} &= -\frac{\partial u}{\partial\delta_j} = b_{ij}^2v_iv_j\sin(\delta_i - \delta_j) \\
        \frac{\partial^2 u}{\partial \delta_i^2} &= \frac{\partial^2 u}{\partial \delta_j^2} = -\frac{\partial^2 u}{\partial\delta_i\partial\delta_j} = b_{ij}^2v_iv_j\cos(\delta_i - \delta_j).
    \end{split}
\end{equation} 

In vector form, we obtain more concisely the following expression:
\begin{equation} \label{eqn:vector-form-line-energy-derivatives}
    \begin{split}
        \nabla_{\boldsymbol{z}}u = 
        \begin{bmatrix}
            \nabla_{\boldsymbol{v}}u      \\
            \mathbf{0} \\
            \nabla_{\boldsymbol{\delta}}u \\
            \mathbf{0}
        \end{bmatrix}, 
        H_{\boldsymbol{z}}u = 
                \begin{bmatrix}
                {H}_{\boldsymbol{v}}u & \mathbf{0} & H_{\boldsymbol{v}\boldsymbol{\delta}}u & \mathbf{0} \\
                \mathbf{0} & \mathbf{0} & \mathbf{0} & \mathbf{0} \\
                H_{\boldsymbol{\delta}\mathbf{v}}u & \mathbf{0} & H_{\boldsymbol{\delta}}u & \mathbf{0} \\
                \mathbf{0} & \mathbf{0} & \mathbf{0} & \mathbf{0}
            \end{bmatrix},
    \end{split}
\end{equation} where $\mathbf{0}$ denotes the vector/matrix of all $0$'s with appropriate dimensions. The notation $H_{\boldsymbol{z}_1\boldsymbol{z}_2}$ is understood as the matrix of mixed (second-order) derivatives involving components of arbitrary vectors $\boldsymbol{z}_1, \boldsymbol{z}_2$. From the vector form of the stochastic differential equations (\ref{eqn:multimachine-model}), we have
\begin{equation} \label{eqn:vector-form-drift-and-diffusion}
    \begin{split}
        \boldsymbol{\mu}_t &= 
        \begin{bmatrix}
            \mathbf{0}\\
            \dot{\boldsymbol{\omega}}(t) \\
            \dot{\boldsymbol{\delta}}(t)\\
            \dot{\boldsymbol{\eta}}(t)-\alpha\sqrt{2\theta}\boldsymbol{C}\cdot\frac{d\boldsymbol{W}(t)}{dt}
        \end{bmatrix},  \\
        \boldsymbol{\sigma}_t &= 
        \begin{bmatrix}
            \mathbf{0} & & & \\
            & \mathbf{0} & & \\
            & & \mathbf{0} & \\
            & & & \alpha\sqrt{2\theta}\cdot\boldsymbol{C}
        \end{bmatrix}
    \end{split}
\end{equation} in correspondence with the notations introduced in (\ref{eqn:reduced-order-pdf-equation}). Consequently, we can readily compute the quantities defined formally in (\ref{eqn:reduced-order-drift}) and (\ref{eqn:reduced-order-diffusion}) as the following:
\begin{equation} \label{eqn:final-response-variable-advection-diffusion}
    \begin{split}(\nabla_{\boldsymbol{z}} u)^T\boldsymbol{\mu}_t &= (\nabla_{\boldsymbol{\delta}}u)^Td\boldsymbol{\delta}(t) \\
        &= b_{ij}^2v_iv_j\sin(\delta_i-\delta_j)(w_i-w_R) 
        \\
        &- b_{ij}^2v_iv_j\sin(\delta_i-\delta_j)(w_j-w_R) 
        \\
        &= b_{ij}^2v_iv_j\sin(\delta_i-\delta_j)(\omega_i-\omega_j);
    \end{split}
\end{equation} and since $\boldsymbol{\sigma}_t^T(H_{\boldsymbol{z}}u)\boldsymbol{\sigma}_t = \mathbf{0}$, $(\nabla_{\boldsymbol{z}} u)^T\boldsymbol{\sigma}_t = \mathbf{0}$, we obtain:
\begin{equation} \label{eqn:final-drift-diffusion-computation}
    \begin{split}
        \mu^u &= (\nabla_{\boldsymbol{z}}u)^T\boldsymbol{\mu}_t + 
        \frac12\text{tr}(\boldsymbol{\sigma}_t^T(H_{\boldsymbol{z}}u)\boldsymbol{\sigma}_t) \\
              &=
        b_{ij}^2v_iv_j\sin(\delta_i-\delta_j)(\omega_i-\omega_j) \\
        \mathcal{D}^u &= (\nabla_{\boldsymbol{z}} u)^T\boldsymbol{\sigma}_t((\nabla_{\boldsymbol{z}} u)^T\boldsymbol{\sigma}_t)^T = 0.
    \end{split}
\end{equation}

Here, the (stochastic) differentials $d\boldsymbol{\omega}(t), d\boldsymbol{\delta}(t), d\boldsymbol{\eta}(t)$ are defined through rewriting the system (\ref{eqn:multimachine-model}) in the It\^o interpretation, consistent with the general definition in (\ref{eqn:general-form-sde}). The above derivations extend trivially to the multiple-lines case (\ref{eqn:m-dim-ropdf}). 

 \vspace{0.3cm}
    \begin{center}
        \scriptsize \framebox{
            \parbox{2.5in}{
                Government License (will be removed at publication):
                The submitted manuscript has been created by UChicago Argonne, LLC,
                Operator of Argonne National Laboratory (``Argonne").  Argonne, a
                U.S. Department of Energy Office of Science laboratory, is operated
                under Contract No. DE-AC02-06CH11357.  The U.S. Government retains for
                itself, and others acting on its behalf, a paid-up nonexclusive,
                irrevocable worldwide license in said article to reproduce, prepare
                derivative works, distribute copies to the public, and perform
                publicly and display publicly, by or on behalf of the Government. The Department of Energy will provide public access to these results of federally sponsored research in accordance with the DOE Public Access Plan. http://energy.gov/downloads/doe-public-access-plan.
            }
        }
        \normalsize
    \end{center}

\end{document}